\documentclass{article}

\usepackage{arxiv}

\usepackage[utf8]{inputenc} 
\usepackage[T1]{fontenc}    
\usepackage{hyperref}       
\usepackage{url}            
\usepackage{booktabs}       
\usepackage{amsfonts,amssymb,amsthm,amsmath,amscd,mathrsfs,latexsym}
\usepackage{nicefrac}       
\usepackage{microtype}      
\usepackage{lipsum}
\usepackage{graphicx}
\graphicspath{ {./images/} }
\newcommand{\indep}{\perp \!\!\! \perp}

\usepackage{amsmath}

\usepackage[capitalize, noabbrev]{cleveref}

\usepackage{tikz} 
\usetikzlibrary{positioning, calc, shapes.geometric, shapes, shapes.multipart, arrows.meta, arrows, decorations.markings, external, trees}
\usetikzlibrary{shapes,decorations,arrows,calc,arrows.meta,fit,positioning}
\tikzset{
	-Latex,auto,node distance =1.5 cm and 1.5 cm,semithick,
	state/.style ={minimum width = 1.5 cm},
	point/.style = {inner sep=0.4cm,fill,node contents={}},
	bidirected/.style={Latex-Latex},
	el/.style = {inner sep=2pt, align=left, sloped}
}
\usepackage{subfig} 
\usepackage{algorithm}
\usepackage{algpseudocode}

\title{A Bayesian model averaging approach for estimating longitudinal treatment effects}

\author{
 David Adenyo \\ 
  Department of Social and Preventive Medicine\\
  Universit\'e Laval\\
   \And
 Mireille E. Schnitzer \\ 
  Faculty of Pharmacy \& Department of Social and Preventive Medicine\\
  Universit\'e de Montr\'eal\\
  Department of Epidemiology, Biostatistics, and Occupational Health \\
  McGill University \\
  \And
 Jason R. Guertin \\ 
  Department of Social and Preventive Medicine\\
  Universit\'e Laval\\
  \And
 Denis Talbot \\ 
  Department of Social and Preventive Medicine\\
  Universit\'e Laval\\
  \texttt{denis.talbot@fmed.ulaval.ca}
}

 \newcommand{\bitem}{\begin{itemize}}
 \newcommand{\eitem}{\end{itemize}}

\usepackage[figuresright]{rotating}
\usepackage{amsmath, amssymb}
\usepackage{booktabs}
\usepackage{natbib}
\usepackage{threeparttable}
\usepackage{multirow}
\usepackage{url}

\begin{document}
\maketitle
\begin{abstract} 
Several methods have been proposed for performing variable selection for the goal of estimating the average treatment effect of a point treatment on an outcome. Much less work has focused on developing methods for performing variable selection when the goal is instead to estimate the causal effect of a longitudinal treatment on an outcome. We introduce a Bayesian model averaging approach to estimate counterfactual means under longitudinal treatment regimes as well as the parameters of longitudinal marginal structural models. The Bayesian model averaging framework provides a principled way to produce valid post-selection inference. We further devise an informative prior distribution that favors the selection of optimal adjustment sets, in the sense that the resulting estimator has the lowest possible variance among all possible adjustment sets. We show through simulation results that our proposed method outperforms fully adjusted estimators and even performs similarly to an oracle estimator in several scenarios. In a real-data illustration, our method yields estimates with markedly reduced standard errors for the effect of cumulative exposure to statins for the prevention of a first cardiovascular disease or death among older adults.
\end{abstract}


\section{Introduction \label{s:intro}}

One of the main challenges in estimating causal effects using observational data is to identify and measure potential confounders. This task is particularly challenging when the goal is to estimate the effect of a time-varying treatment since controlling for both baseline and time-varying variables may be required to satisfy the sequential conditional exchangeability assumption. While these variables are usually identified using domain-specific knowledge and experts' opinions, it may be unknown whether control for some variables is truly needed to satisfy exchangeability. A common strategy is then to adjust for ``everything but the kitchen sink'', wherein all variables suspected to be potential confounders based on subject-matter knowledge are adjusted for. This may result in adjusting for more variables than needed, increasing the variance of the causal effect estimator \citep{rotnitzky2010note,rotnitzky2020efficient}, and potentially exacerbating practical violations of positivity \citep{rudolph2022estimation}, as compared to a more parsimonious adjustment set. In a point treatment setting, previous studies demonstrated that it is better to adjust for covariates that are true confounding covariates (those associated with both treatment and outcome) in addition to those associated only with the outcome, and to avoid adjusting for those associated only with the treatment  \citep[e.g.,][]{rotnitzky2010note, rotnitzky2020efficient}. Similar results have also been produced in the time-varying treatment setting \citep{rotnitzky2020efficient, adenyo2024longitudinal}.

Several statistical approaches have been developed to use the data as a supplement to subject-matter knowledge for selecting adjustment variables when estimating the effect of a treatment measured at a single time-point on an outcome \citep[e.g., see][for reviews]{witte2019covariate, talbot2022generalized}. Many of these approaches have demonstrated their ability to produce causal effect estimators with a reduced variance as compared to a fully adjusted model. 
However, very few variable selection methods have been developed for the goal of omitting variance-inflating variables while estimating the effect of a time-varying treatment on an outcome. In this specific setting, a collaborative longitudinal targeted maximum likelihood estimation (C-LTMLE) approach has been proposed by \citep{schnitzer2020data}. This approach is based on LTMLE, a doubly robust estimator that requires estimating both the conditional treatment probabilities and a sequence of conditional nested outcome expectations but is consistent if either is consistently estimated. C-LTMLE involves stepwise variable selection in the treatment models, with selection performed conditional on preliminary estimates of the nested outcome expectations. Results from simulation studies conducted in a point treatment setting suggest that further variance reduction may sometimes be achieved by performing variable selection in both the treatment and outcome models \citep{talbot2022generalized}. A longitudinal outcome adaptive lasso has also been proposed by \citep{schnitzer2024adaptive}. However, producing valid post-selection inferences remains a challenge for both of these estimators \citep{schnitzer2024adaptive, schnitzer2020data}.

Building on previous work \citep{talbot2022generalized}, we introduce a longitudinal Bayesian model averaging approach for selecting adjustment variables, which we call the Longitudinal Generalized Bayesian Causal Effect Estimation (LGBCEE) algorithm. Our proposed approach is based on the doubly robust LTMLE, performs variable selection for both the treatment and outcome models, and leverages the Bayesian framework to account for model selection in order to produce valid post-selection inferences in a principled way. 

\section{Data structure and review of marginal structural models\label{s:structure}}

We consider a study with $T+1$ time-points where we are interested in the effect of a time-varying treatment, $A_t, t=0,...,T$, on an end-of-study outcome, $Y$. We denote $\bar{A}_{t}=\{A_{0},{A}_{1},...,{A}_{t}\}$, the history of the binary treatment (or exposure) up to time $t$. Let $X_t = \{X_{t1},...,X_{tN_t}\}$ be a vector of observed time-varying potential confounders at time $t$ and let $\bar{X}_{t}=\{X_{0},{X}_{1},...,{X}_{t}\}$ be the potential confounder history up to time $t$. We assume the following observed data time-ordering: $\mbox{O} = \{X_0, A_0, X_1, A_1, ..., X_T, A_T, Y\}$. Let $\mathcal{A}$ be the set of all possible longitudinal treatment regimes  and let $\bar{a}_{T}=(a_0,a_1,...,a_T)$ be a given element in that set. Let $Y^{\bar{a}_{T}}$ be the counterfactual outcome that would have been observed if, possibly contrary to fact, $\bar{A}_{T}=\bar{a}_{T}$.

A common framework for representing the effect of time-varying treatments is marginal structural models (MSMs) \citep[][]{robins2000marginal}. MSMs represent a parametrized model for the marginal mean of the counterfactual outcome in terms of the treatment regime, e.g. $   \mathbb{E}(Y^{\bar{a}_{T}}) = \phi(\bar{a}_{T},\beta)$. 
Several parameterizations are possible for the function $\phi$. For simplicity and without loss of generality, we focus on 
\begin{align}
   \mathbb{E}(Y^{\bar{a}_{T}}) = \beta_0+\beta_{1}\sum_{t=0}^T a_t \label{eq:workingMSM}
\end{align}
in this article. We let $\Delta$ denote a vector of the causal quantities of interest, which are functions of the parameters of the MSM. For example, $\Delta$ could be the parameters of the MSM ($\beta_0,\beta_1$), a counterfactual expectation (e.g., $\Delta = \mathbb{E}(Y^{\bar{a}_{T} = \bar{1}}) = \beta_0 + ({T+1}) \beta_1$) or a contrast between counterfactual expectations (e.g., $\Delta = \mathbb{E}(Y^{\bar{a}_{T} = \bar{1}}) - \mathbb{E}(Y^{\bar{a}_{T} = \bar{0}}) = ({T+1}) \beta_1$).

In order to imbue a causal interpretation to MSM parameters based on the observed data, we make the following causal assumptions: no interference, conditional exchangeability, positivity and consistency \citep{hernan2020what}. No interference means that the treatment of one individual does not affect the counterfactual outcomes of the others. Conditional exchangeability 
$Y^{\bar{a}_{T}} \indep A_t|\bar{A}_{t-1}, \bar{X}_{t}, \ t=0,...,T$, with $\bar{A}_{-1}\equiv \varnothing$, is  the absence of unmeasured confounding factors. Positivity means that $0< P(A_t=1|\bar{A}_{t-1},\bar{X}_{t})< 1$ almost surely at every time-point $t$. Consistency means that, for any individual, if $\bar{A}_{T}=\bar{a}_{T}$, then $Y=Y^{\bar{a}_{T}}$. Under these assumptions, counterfactual expectations are nonparametrically identified as 
\begin{align}
    \mathbb{E}(Y^{\bar{a}_{T}}) = \mathbb{E}\left(\mathbb{E}\left[...\mathbb{E}\left\{\mathbb{E}\left(Y|\bar{A}_T = \bar{a}_T,\bar{X}_T\right)|\bar{A}_{t-1} =\bar{a}_{t-1},\bar{X}_{t-1}\right\}...|A_0 = a_0,X_0\right]\right) \label{eq:id1}
\end{align}
\citep[see e.g., ][]{bang2005doubly}. Letting $Q^{\bar{a}_T}_{T+1} = Y$ for all $\bar{a}_T$ and iteratively defining $Q_t^{\bar{a}_T} = \mathbb{E}\left(Q^{\bar{a}_T}_{t+1}|\bar{A}_t = \bar{a}_t,\bar{X}_t\right)$ for $t = 0, ..., T$, expression (\ref{eq:id1}) can be written as $\mathbb{E}(Q_{0}^{\bar{a}_T})$. Expression (\ref{eq:id1}) can sometimes be simplified by exploiting independencies. Let $Z_t \subseteq \{\bar{A}_{t-1}, \bar{X}_t\}$ such that $Q_t^{\bar{a}_T} \indep \{\bar{A}_{t-1}, \bar{X}_t\}\backslash Z_t \mid Z_t$, then 
 the identification formula becomes
\begin{align}
    \mathbb{E}(Y^{\bar{a}_{T}}) = \mathbb{E}\left(\mathbb{E}\left[...\mathbb{E}\left\{\mathbb{E}\left(Y|A_t = a_T,Z_T(\bar{a}_{t-1})\right)|A_{t-1} = a_{t-1},Z_{t-1}(\bar{a}_{t-2})\right\}...|A_0 = a_0,Z_0\right]\right), \label{eq:id2}
\end{align}
where the notation $Z_k(\bar{a}_{k-1})$ is used to emphasize that if some or all past treatments are included in $Z_k$, then they are held fixed in (\ref{eq:id2}). Exploiting such independencies is a key to reducing the variance of causal effect estimators in the longitudinal treatment setting \citep[][]{adenyo2024longitudinal}.

To simplify the presentation moving forward, we consider the case with two time-points ($T = 1$). See the Online Supplement for an extension.

\subsection{The Pooled and Stratified Longitudinal Maximum Likelihood Estimator (LTMLE)\label{s:PLTMLE}}

As mentioned previously, our proposed approach employs LTMLE to estimate the parameters of the MSM (\ref{eq:workingMSM}) under the preceding causal assumptions. We use both stratified and pooled LTMLE estimators \citep{schnitzer2014modeling, van2012targeted, petersen2014targeted}. We now present an example of the pooled LTMLE algorithm for the working MSM we consider based on \citep{petersen2014targeted}. In the following, we assume that $Y$ is binary or continuous and rescaled to be bounded between 0 and 1.
\begin{enumerate}
\item \textit{Step 1}. Estimate $f_{1}(Z_{1})={P}(A_1|Z_1)$ and $f_{0}(Z_{0})={P}(A_0|Z_{0})$ by regressing $A_1$ on $Z_1$, and  $A_0$  on $Z_{0}$, respectively. Denote the corresponding estimators $f_{1,n}$ and $f_{0,n}$ respectively.

\item \textit{Step 2}. Generate a vector $\bar{Q}_{1,n}^{*}(Z_1)$ of length $4 \times n$
\begin{enumerate}
    \item Regress $Y$ on $A_1$, $Z_1$ and generate predicted values $Q^{\bar{a}_1}_{1,n}(Z_1)$ for each individual for each of the four possible treatment regimes $\bar{a}_1 \in \mathcal{A}$. Stack the predictions to create a vector $\bar{Q}_{1,n}(Z_1)$ of length $4\times n$.
    \item For each subject, create a vector consisting of four copies of their observed outcome $Y_i$ (one for each treatment regime) and stack these to create a $n \times 4$ vector denoted $\bar{Q}^{*}_{2,n}$.
    \item Create a covariate matrix with $4$ rows and $2$ columns for each subject by taking partial derivatives of the working MSM (i.e., Equation \ref{eq:workingMSM}) with respect to $\beta_0$ and $\beta_1$:  $(1, (a_0+a_1))$ for each $\bar{a}_{1} \in \mathcal{A}$. The corresponding matrix for all subjects therefore has $4 \times n$ rows and $2$ columns.
    \item Fit a pooled logistic regression of $\bar{Q}_{2,n}^{*}$ on the previous covariate matrix, suppressing the intercept and using  $\text{logit}(\bar{Q}_{1,n}(Z_1))$ as offset and $\frac{I(\bar{A}_1 = \bar{a}_1)}{f_{0,n}f_{1,n}}$ as weights. This estimates a multivariate parameter $\varepsilon_1$, corresponding to the coefficients of the two covariate matrix columns. Denote these estimators $\varepsilon_{1,n} = (\varepsilon_{1,n}^{\beta_0},\varepsilon_{1,n}^{\beta_1})$.
    \item Generate $\bar{Q}_{1,n}^*(Z_1)$ by evaluating, for each subject and each $\bar{a}_{1} \in \mathcal{A}$,
    $\mbox{expit}[\mbox{logit}\{{Q}_{1,n}^{\bar{a}_1}(Z_1)\}+1\times \varepsilon_{1,n}^{\beta_0}+(a_0+a_1)\times \varepsilon_{1,n}^{\beta_1}].$
\end{enumerate}
\item \textit{Step 3}. Generate a vector $\bar{Q}_{0,n}^*(Z_0)$ of length $n \times 4$.
\begin{enumerate}
    \item Fit a logistic regression of $\bar{Q}_{1,n}^*(Z_1)$ on $A_0$, $Z_0$ and generate predictions ${Q}_{0,n}^{\bar{a}_1}(Z_0)$ for each subject and each $\bar{a}_1 \in \mathcal{A}$. Denote these stacked predictions $\bar{Q}_{0,n}(Z_0)$.
    \item Fit a pooled logistic regression of $\bar{Q}_{1,n}^{*}(Z_1)$ on the covariate matrix of step 2c, suppressing the intercept and using  $\text{logit}(\bar{Q}_{0,n}(Z_0))$ as offset and $\frac{I(A_0 = a_0)}{f_{0,n}}$ as weights. This estimates $\varepsilon_0$, with estimator $\varepsilon_{0,n}$.
    \item Generate $\bar{Q}_{0,n}^*(Z_0)$ by evaluating, for each subject and each $\bar{a}_{1} \in \mathcal{A}$,
    $\mbox{expit}[\mbox{logit}\{{Q}_{0,n}^{\bar{a}_1}(Z_0)\}+1\times \varepsilon_{0,n}^{\beta_0}+(a_0+a_1)\times \varepsilon_{0,n}^{\beta_1}].$
\end{enumerate}
\item \textit{Step 4}. If $Y$ was rescaled, apply the inverse transformation to $\bar{Q}_{0,n}^{*}(Z_0)$ to return to the original scale. Estimate the parameters of the working MSM by fitting a linear regression of $\bar{Q}_{0,n}^{*}(Z_0)$ on $a_0+a_1$. Estimate the $\Delta$ using the MSM parameter estimates.  
\end{enumerate}
                    
Stratified LTMLE proceeds similarly, except that a separate $\varepsilon$ model is fitted for each time $t$ and each regime of interest. 

Using LTMLE for estimating the parameters of an MSM has several advantages. First, if the covariates used to produce the estimates $\{f_{1,n}(Z_{1}),f_{0,n}(Z_{0})\}$ and $\{\bar{Q}_{1,n}^*(Z_{1}),\bar{Q}_{0,n}^{*}(Z_{0})\}$ satisfy the conditional exchangeability assumption and the estimators are consistent for $\{f_{1}(Z_{1})$, $f_{0}(Z_{0})\}$ and $\{Q_1^{\bar{a}_1}, Q_0^{\bar{a}_1}\}$ for all $\bar{a}_1 \in \mathcal{A}$, respectively, then the resulting estimator of $\Delta$ attains the semi-parametric efficiency bound, i.e., it has the smallest asymptotic variance among all semi-parametric estimators of $\Delta$ that condition on the same $Z_1$, $Z_0$ \citep[][]{petersen2014targeted}. In addition, if either $f_{1}(Z_{1})$ and $f_{0}(Z_{0})$, or $Q_{1}^{\bar{a}_{1}}$ and $Q_{0}^{\bar{a}_{1}}$ are consistently estimated, then the LTMLE estimator is consistent for the target parameter. In other words, the estimator for $\Delta$ is doubly robust. LTMLE also enables the use of flexible learning algorithms to estimate nuisance functions, although the ability to produce valid inferences then relies on convergence rate conditions, as well as smoothness (e.g., Donsker class) conditions or the use of sample splitting techniques \citep[][]{aa, balzer2023invited, chernozhukov2018double, zheng2010asymptotic}. In our proposed approach, parametric regression models are instead used, thus relying on correct model specification to avoid potential bias.

\section{Variable selection objective in a time-varying treatment setting \label{s:selection.principle}}

The LGBCEE algorithm we propose leverages information from both the treatment-covariate associations and outcome-covariate associations to favor adequate control of confounding with reduced variance in finite samples. As previously observed, focusing only on the outcome-covariate associations may result in not controlling for important confounders that are weakly associated with outcome but strongly associated with treatment \citep[e.g.][]{wang2012bayesian, talbot2015bayesian}. Based on \citep{rotnitzky2020efficient} and \citep{adenyo2024longitudinal}, we thus formulate the following desiderata. First, to control for confounding, a variable should be included in $Z_t$ if it is associated with $A_t$ conditional on the other variables in $Z_t$ and also associated with $Q_{t+1}^{\bar{a}_{1}}$ conditional on $A_t$ and the other variables in $Z_t$. To minimize the asymptotic variance of the estimator, a variable should also be included in $Z_t$ if it is not associated with $A_t$ conditional on the other variables in $Z_t$ but it is associated with $Q_{t+1}^{\bar{a}_{1}}$ conditional on $A_t$ and the other variables in $Z_t$. Opposingly, a variable should be excluded from $Z_t$ if it is associated with $A_t$ conditional on the other variables in $Z_t$ but not associated with $Q_{t+1}^{\bar{a}_{1}}$ conditional on $A_t$ and the other variables in $Z_t$. The inclusion/exclusion of variables that are neither associated with $A_t$ conditional on the other variables in $Z_t$, nor with  $Q_{t+1}^{\bar{a}_{1}}$ conditional on $A_t$ and the other variables in $Z_t$ does not affect the asymptotic variance of the estimator, but excluding these variables can be beneficial to reduce the finite sample variance. 

\section{The LGBCEE algorithm\label{s:LGBCEE}}

\subsection{Overview in the two time-point case\label{rat data}}
To facilitate the presentation, we introduce some additional notation. Let ${U}_{1}=\{U_{11},...,U_{1N_0},U_{1(N_0+1)},U_{1(N_0+2)},...,U_{1(N_0+N_1+1)}\}$
where $U_{1n_0} = X_{0n_0}$ for $n_0 = 1,...,N_0$; $U_{1(N_0+1)} = A_{0}$; $U_{1(N_0+n_1+1)} = X_{1n_1}$ for $n_1 = 1,...,N_1$; and ${U}_{0}=\{U_{01},...,U_{0N_0}\}$, where $U_{0n_0} = X_{0n_0}$ for $n_0 = 1,...,N_0$. In other words, $U_{1}$ and $U_0$ represent all the variables available for adjustment at time-points 1 and 0, respectively, including previous treatments. 

The LGBCEE algorithm entails a Bayesian averaging of the LTMLE estimates over all combinations of potential adjustment sets at each time-point. Let $\alpha^{Y}=(\alpha_{11}^{Y},...,\alpha_{1N_0}^{Y},\alpha_{1(N_0+1)}^{Y},\alpha_{1(N_0+2)}^{Y},... ,\alpha_{1(N_0+N_1+1)}^{Y})$ be a vector of dimension $M_1=N_0+N_1+1$ for the inclusion of ${U}_{1}=\{X_0,A_0,X_{1}\}$ in the model for $Y$, where a component of $\alpha^{Y}$ is equal to $1$ if the variable associated with it is included in the model, and $0$ otherwise. Similarly, $\alpha^{A_1}$ is a vector of dimension $M_1$ for the inclusion of $U_1$ in the model for $A_1$, $\alpha^{Q_1}=(\alpha_{01}^{Q_1},...,\alpha_{0N_0}^{Q_1})$ is a vector of dimension $M_0 = N_0$ for the inclusion of the covariates $U_0$ in the model for $Q_1^{\bar{a}_1}$, and $\alpha^{A_0}$ is a vector of dimension $M_0$ for the inclusion of covariates $U_0$ in the model for $A_0$. Note that the same variables are included in the exposure and outcome models at a given time-point when computing an LTMLE estimate in LGBCEE (i.e., $\alpha^{A_1} = \alpha^{Y}$ and $\alpha^{A_0} = \alpha^{Q_1}$). The rationale for this choice is based on the sequential conditional exchangeability assumption where a given adjustment set is used at each time-point. We nonetheless introduce separate notation for the treatment and outcome models because, as will be seen shortly, LGBCEE requires computing the posterior probabilities of the possible treatment models before computing the LTMLE estimates. Let $\mathcal{A}_1$ be the set of all possible values of $\alpha^{Y}$ and $\alpha^{A_1}$, and $\mathcal{A}_0$ be the set of all possible values of $\alpha^{Q_1}$ and $\alpha^{A_0}$. 

We define $\hat{\Delta}^{\alpha^{Y},\alpha^{Q_1}}$ to be an LTMLE estimate of $\Delta$ obtained by adjusting for the variables indicated by $\alpha^{Y} = \alpha^{A_1}$ for estimating $Q_{1}^{\bar{a}_{1}}$ and $f_{1}^{\bar{a}_{1}}(\cdot)$, and for the variables indicated by $\alpha^{Q_1} = \alpha^{A_0}$ for estimating $Q_{0}^{\bar{a}_{1}}$ and $f_{0}^{{a}_{0}}(\cdot)$. LGBCEE estimates $\Delta$ as a model-averaged posterior mean 
\[\hat{\Delta}\equiv {\mathbb{E}}[\Delta|\mathbf{\mbox{O}}]=\sum_{(\alpha^{Y},\alpha^{Q_1})\in \mathcal{A}_1\times \mathcal{A}_0}
\hat{\Delta}^{\alpha^{Y},\alpha^{Q_1}}{P}(\alpha^{Y},\alpha^{Q_1}|\mbox{O}),\] 
where $\hat{\Delta}^{\alpha^{Y},\alpha^{Q_1}}\equiv {\mathbb{E}}[\Delta|\mathbf{\mbox{O}},\alpha^{Y},\alpha^{Q_1}]$ is the posterior mean of $\Delta$ after adjustment for the variables indicated by $\alpha^{Y}$ and $\alpha^{Q_1}$ and ${P}(\alpha^{Y},\alpha^{Q_1}|\mbox{O})={P}(\alpha^{Y}|\mbox{O}){P}(\alpha^{Q_1}|\alpha^{Y},\mbox{O})$ is the weight attributed to each estimate, i.e., the posterior probability of the adjustment set ($\alpha^{Y}$, $\alpha^{Q_1}$).

The specific models used within the LTMLE estimator are as follows:
\begin{align}
g[{\mathbb{E}}(Y|\alpha^{Y},{A}_{1},{U}_{1})]&=\delta_{1}^{\alpha^{Y}}+\beta^{\alpha^{Y}}A_1+\sum_{m_1=1}^{M_1}\alpha_{1m_1}^{Y}\delta_{1m_1}^{\alpha^{Y}}U_{1m_1}, \label{eq1b}   \\
g[{\mathbb{E}}(Q_1^{\bar{a}_1}|\alpha^{Y},\alpha^{Q_1}, {A}_{0},{U}_{0})]&=\delta_{0}^{\alpha^{Q_1}}+\beta^{\alpha^{Q_1}}A_0+\sum_{m_0=1}^{M_0}\alpha_{0m_0}^{Q_1} \delta_{0m_0}^{\alpha_{0}^{Q_1}}U_{0m_0}, \label{eq2b} \\
\text{logit}\left[ P(A_1=1|\alpha^{A_1},{U}_{1})\right] &=\delta_{1}^{\alpha^{A_1}}+\sum_{m_1=1}^{M_1}\alpha_{1m_1}^{A_1} \delta_{1m_1}^{\alpha^{A_1}}U_{1m_1},\label{eq3b} \\
\text{logit}\left[ P(A_0=1|\alpha^{A_0},{U}_{0})\right] &=\delta_{0}^{\alpha^{A_0}}+\sum_{m_0=1}^{M_0}\alpha_{0m_0}^{A_0} \delta_{0m_0}^{\alpha^{A_0}}U_{0m_0},\label{eq4b}
\end{align}
where $g$ is a known link function. Note that the model in Equation (\ref{eq2b}) is conditional on $\alpha^{Y}$ because the estimate of $Q_1^{\bar{a}_1}$ is calculated from the output of model (\ref{eq1b}) in the LTMLE algorithm.

As will be seen shortly, we use the prior distribution ${P}(\alpha^{Y}|A_1,U_1){P}(\alpha^{Q_1}|\alpha^{Y},A_0,U_0)$ to achieve the desired variable selection objective introduced in Section \ref{s:selection.principle}. Using standard Bayesian calculations, the posterior distribution of the adjustment set $(\alpha^{Y},\alpha^{Q_1})$ is ${P}(\alpha^{Y},\alpha^{Q_1}|\mbox{O})\propto P(Y|\alpha^{Y},{A}_1,{U}_1)P(Q_1^{\bar{a}_1}|\alpha^{Y},\alpha^{Q_1}, A_0,U_0){P}(\alpha^{Y}|A_1,U_1){P}(\alpha^{Q_1}|\alpha^{Y},A_0,U_0)$.

\subsection{Prior distribution construction \label{s:prior.construction}}
The first step of the construction of the prior distribution is to identify the determinants of the treatment at each time-point.  
To do so, we use the treatment models \eqref{eq3b} and \eqref{eq4b} and calculate the posterior distributions ${P}(\alpha^{A_1}|{A}_{1},{U}_{1})\propto {P}(A_1|\alpha^{A_1},{U}_{1}){P}(\alpha^{A_1})$ and 
${P}(\alpha^{A_0}|A_0,U_0)\propto {P}(A_0|\alpha^{A_0},U_0){P}(\alpha^{A_0})$. 
Classical Bayesian model averaging results ensure that the posterior mass concentrates on the true treatment models (i.e., the models that only include the true determinants of the treatment at each time-point) if models \eqref{eq3b} and \eqref{eq4b} are correctly specified \citep{haughton1988choice, wasserman2000bayesian}. 

Similar to \citep{wang2012bayesian}, we make the following assumptions to implement LGBCEE: (\textbf{A1}): $\alpha^{Q_{t+1}}$ is independent of (${A}_{t},{U}_{t}$) given ($\alpha^{Q_{t+2}},\alpha^{A_t}$) and (\textbf{A2}): $\alpha^{A_t}$ is independent of $\alpha^{Q_{t+2}}$ given (${A}_{t},{U}_{t}$). Using these assumptions and the above information, we define the prior distribution of LGBCEE at each time-point as follows:
\begin{eqnarray*}
    {P}(\alpha^{Q_{t+1}}|\alpha^{Q_{t+2}},{A}_{t},{U}_{t})
    &=& \sum_{\alpha^{A_t}\in \mathcal{A}_{t}} {P}(\alpha^{Q_{t+1}}|\alpha^{Q_{t+2}},\alpha^{A_t},{A}_{t},{U}_{t}){P}(\alpha^{A_t}|\alpha^{Q_{t+2}},{A}_{t},{U}_{t}) \\
    &\overset{\text{using (\textbf{A1}),(\textbf{A2})}}{=}&
     \sum_{\alpha^{A_t}\in \mathcal{A}_{t}} {P}(\alpha^{Q_{t+1}}|\alpha^{Q_{t+2}},\alpha^{A_t}){P}(\alpha^{A_t}|{A}_{t},{U}_{t}),  \ t=0,1,
\end{eqnarray*}
where $\alpha^{Q_{3}}=\emptyset$, $\alpha^{Q_{2}}\equiv \alpha^{Y}$ and 
\begin{equation}
{P}(\alpha^{Q_{t+1}}|\alpha^{Q_{t+2}},\alpha^{A_t})\propto \prod_{m_{t}=1}^{M_t} {P}(\alpha_{tm_{t}}^{Q_{t+1}}|\alpha^{Q_{t+1}}_{t,-m_t},\alpha^{Q_{t+2}},\alpha_{tm_{t}}^{A_t}), \label{eq.9}
\end{equation}
\begin{equation}
 {P}(\alpha_{tm_{t}}^{Q_{t+1}}=1|\alpha^{Q_{t+1}}_{t,-m_t},\alpha^{Q_{t+2}},\alpha_{tm_{t}}^{A_t}=0)={P}(\alpha_{tm_{t}}^{Q_{t+1}}=0|\alpha^{Q_{t+1}}_{t,-m_t},\alpha^{Q_{t+2}},\alpha_{tm_{t}}^{A_t}=0)=\frac{1}{2},\label{eq.10}
\end{equation}
\begin{equation}
  {P}(\alpha_{tm_{t}}^{Q_{t+1}}=1|\alpha^{Q_{t+1}}_{t,-m_t},\alpha^{Q_{t+2}},\alpha_{tm_{t}}^{A_t}=1)=\frac{\omega_{tm_{t}}^{\alpha^{Q_{t+1}}}}{\omega_{tm_{t}}^{\alpha^{Q_{t+1}}}+1}, \label{eq.Prior3}
\end{equation}
where $\omega_{tm_{t}}^{\alpha^{Q_{t+1}}}$ is proportional to the conditional association between $U_{tm_t}$ and $Q_{t+1}^{\bar{a}_1}$, given the other variables in the outcome model, and $\alpha^{Q_{t+1}}_{t,-m_t}$ denotes the vector $\alpha^{Q_{t+1}}$ without the element related to $U_{tm_t}$. The quantity $\omega_{tm_{t}}^{\alpha^{Q_{t+1}}}$ will be formally defined shortly (in (\ref{eq.Omega})). Note that the prior for $\alpha^{Q_1}$ is conditional on $\alpha^{Q_2}$ because the estimate of $Q_1^{\bar{a}_1}$ is built from the output of model $\alpha^{Q_2}$. In other words, if a potential confounder is not included in the model for $A_t$ (i.e., if $\alpha_{tm_{t}}^{A_{t}}=0$), then it has the same prior probability of being included or excluded in the model for $Q_{t+1}^{\bar{a}_1}$, regardless of its association with $Q_{t+1}^{\bar{a}_1}$. On the other hand, if a potential confounder is included in the model for $A_t$ (i.e., $\alpha_{tm_{t}}^{A_{t}}=1$), then its prior probability of being included in the model for $Q_{t+1}^{\bar{a}_1}$ depends on its association with $Q_{t+1}^{\bar{a}_1}$. In this case, the greater the association between a potential confounder and $Q_{t+1}^{\bar{a}_1}$ conditional on the other variables in the model for $Q_{t+1}^{\bar{a}_1}$, the greater its prior probability of being included in the model for $Q_{t+1}^{\bar{a}_1}$. 

To define $\omega_{tm_{t}}^{\alpha^{Q_{t+1}}}$, denote by $\tilde{\delta}_{tm_t}^{\alpha^{Q_{t+1}}}, t =0,1$ the regression coefficient for $U_{tm_t}$ that measures the association between $U_{tm_t}$ and $Q_{t+1}^{\bar{a}_1}$ conditional on the other variables in $\alpha^{Q_{t+1}}$. Specifically, 
$\tilde{\delta}_{tm_t}^{\alpha^{Q_{t+1}}}={\delta}_{tm_t}^{\alpha^{Q_{t+1}}}$ if 
$U_{tm_t}$ is included in $\alpha_{tm_t}^{Q_{t+1}}$, and $\tilde{\delta}_{tm_t}^{\alpha^{Q_{t+1}}}$ is the regression coefficient for $U_{tm_t}$ in the outcome model that includes the same covariates as $\alpha^{Q_{t+1}}$, but additionally includes $U_{tm_t}$ otherwise. We denote
\begin{equation}
 \omega_{tm_{t}}^{\alpha^{Q_{t+1}}}=\omega\times \left(\tilde{\delta}_{tm_t}^{\alpha^{Q_{t+1}}}\frac{\sigma_{U_{tm_t}}}{\sigma_{Q_{t+1}}}\right)^{2},  \label{eq.Omega} 
\end{equation} 
where $0\leq\omega \leq \infty$ is a user-defined hyperparameter,
which determines how heavily weighted the prior distribution is, $\sigma_{U_{tm_t}}$ and $\sigma_{Q_{t+1}}$
are the standard deviations of $U_{tm_t}$ and $Q_{t+1}^{\bar{a}_1}$, respectively, and $\frac{\sigma_{U_{tm_t}}}{\sigma_{Q_{t+1}}}$ is used to standardize the regression coefficient. In the case where $Q_{t+1}^{\bar{a}_1}$ is binary, we set $\sigma_{Q_{t+1}}=1$ in Equation (\ref{eq.Omega}). In practice, the values of $\tilde{\delta}_{tm_t}^{\alpha^{Q_{t+1}}}$, $\sigma_{U_{tm_t}}$ and $\sigma_{Q_{t+1}}$ are generally unknown and are estimated by their maximum likelihood estimator $\hat{\tilde{\delta}}_{tm_t}^{\alpha^{Q_{t+1}}}$, $s_{U_{tm_t}}$ and $s_{Q_{t+1}}$, respectively, giving an empirical Bayes flavor to the prior distribution. Our proposed implementation described below captures the uncertainty in the parameters of the prior distribution, resulting in adequate coverage of the credible intervals. Since the regression coefficients $\tilde{\delta}_{tm_t}^{\alpha^{Q_{t+1}}}$ are estimated, it is also desirable for the prior distribution to be more heavily weighted when more precise estimates of $\tilde{\delta}_{tm_t}^{\alpha^{Q_{t+1}}}$ are available. This goal is achieved by setting $\omega=c\times n^{b}$, where $0<b<1$ and $c$ is a constant. See \citep{talbot2015bayesian} and \citep{talbot2022generalized} for a more detailed argument.

\section{Implementation}

In this section, we provide a general overview of the practical implementation of LGBCEE. A worked out example using simulated data is available in the Online Supplement. Note that similar to other implementations of Bayesian model averaging, such as that in the BMA package in R, we do not explicitly specify prior distributions for the regression parameters and instead implicitly assume vague priors. 

\textit{Step 1}. Determine the posterior distribution of the treatment models at each time-point, ${P}(\alpha^{A_t}|{A}_{t},{U}_{t})\propto P(A_t|\alpha^{A_t},{U}_{t}){P}(\alpha^{A_t})$, $t=0,1$ where ${P}(\alpha^{A_t})$ is a user-supplied prior distribution for the treatment model. We use ${P}(\alpha^{A_t})\propto 1$. 

\textit{Step 2}. Explore the outcome model space $\alpha^{Y}$ 
using a Markov chain Monte Carlo model composition algorithm \citep{madigan1995bayesian}. The algorithm starts with an initial model. At each step, a candidate model (say, $\alpha_{1}^{Y}$) is proposed by adding or removing one covariate $U_{1m_1^{'}}$ in the current model (say, $\alpha_{2}^{Y}$). The Metropolis-Hastings criterion based on the ratio between the posterior probability of the candidate model and the posterior probability of the current model is then used to decide whether or not to accept this candidate model:
\[
\frac{P(\alpha_{1}^{Y}|\mbox{O})}{P(\alpha_{2}^{Y}|\mbox{O})}=
\frac{P(Y|\alpha_{1}^{Y},{A}_1,{U}_1)}{ P(Y|\alpha_{2}^{Y},{A}_1,{U}_1)}\times 
\frac{\sum_{\alpha^{A_1}}{P}(\alpha_{1}^{Y}|\alpha^{A_1}){P}(\alpha^{A_1}|{U}_1,{A}_1)}{\sum_{\alpha^{A_1}}{P}(\alpha_{2}^{Y}|\alpha^{A_1}){P}(\alpha^{A_1}|{U}_1,{A}_1)}
\]
where ${P}(\alpha_{1}^{Y}|\alpha^{A_1})$ and ${P}(\alpha_{2}^{Y}|\alpha^{A_1})$ can be calculated based on equations (\ref{eq.9}), (\ref{eq.10}), (\ref{eq.Prior3}) and $\frac{P(Y|\alpha_{1}^{Y},{A}_1,{U}_1)}{ P(Y|\alpha_{2}^{Y},{A}_1,{U}_1)}
$ can be approximated by a BIC approximation $\mbox{exp}\left\lbrace
\frac{1}{2}(BIC(\alpha_{1}^{Y})-BIC(\alpha_{2}^{Y}))
\right\rbrace$ \citep{Clyde2003}, where the BIC is calculated based on model (\ref{eq1b}). 
Similar to \citep{talbot2022generalized}, we employ the following heuristic approximation to reduce the computational burden:
\begin{align*}
    & \frac{\sum_{\alpha^{A_1}} \prod_{m_{1}=1}^{M_1} {P}(\alpha_{1,1m_{1}}^{Y}|\alpha_{1,1,-m_{1}}^{Y},\alpha_{1m_{1}}^{A_1}){P}(\alpha^{A_1}|{U}_1,{A}_{1})}{\sum_{\alpha^{A_1}} \prod_{m_{1}=1}^{M_1} {P}(\alpha_{2,1m_{1}}^{Y}|\alpha_{2,1,-m_{1}}^{Y},\alpha_{1m_{1}}^{A_1}){P}(\alpha^{A_1}|{U}_1,{A}_{1})}
     \\ & \qquad \qquad \approx
   \frac{\sum_{\alpha^{A_1}} {P}(\alpha_{1,1m_1^{'}}^{Y}|\alpha_{1,1,-m_1^{'}}^{Y},\alpha_{1m_1^{'}}^{A_1}){P}(\alpha^{A_1}|{U}_1,{A}_{1})}{\sum_{\alpha^{A_1}}{P}(\alpha_{2,1m_1^{'}}^{Y}|\alpha_{2,1,-m_1^{'}}^{Y},\alpha_{1m_1^{'}}^{A_1}){P}(\alpha^{A_1}|{U}_1,{A}_{1})}.
\end{align*}
where $\alpha_{m_1^{'}}^{Y}$ denote the coefficient associated with 
the covariate $U_{1m_1^{'}}$ randomly selected to be added or removed.   

\textit{Step 3}. 
For each $\alpha^{Y}$ explored, calculate $\bar{Q}_{1,n}^*$ and compute the marginal likelihood $P(Q_{1}^{\bar{a}_1}|\alpha^{Y},\alpha^{Q_{1}},A_0,U_0)$ using the BIC approximation under model (\ref{eq2b}).

\textit{Step 4}. 
For each $Q_{1}^{\bar{a}_1}$ and conditionally on outcome model $\alpha^{Y}$, use the Markov chain Monte Carlo model composition algorithm to explore the pseudo-outcome model space $\alpha^{Q_{1}}|\alpha^{Y}$.

\textit{Step 5}. For each ($\alpha^{Y}$, $\alpha^{Q_{1}}$), compute $\hat{\Delta}^{\alpha^{Y},\alpha^{{Q_{1}}}}$, $\widehat{\mbox{Var}}(\hat{\Delta}^{\alpha^{Y},\alpha^{{Q_{1}}}})$ and $P(\alpha^{Y},\alpha^{Q_{1}}|\mbox{O})$. The variance of $\hat{\Delta}^{\alpha^{Y},\alpha^{{Q_{1}}}}$ is estimated as the variance of the empirical efficient influence function divided by $n$ \citep{petersen2014targeted}. We have found in our simulation studies that PLTMLE overestimates the true variance when the treatment $A_0$ is not included in models $\alpha^Y$. We thus propose to use PLTMLE to estimate $\hat{\Delta}^{\alpha^{Y},\alpha^{{Q_{1}}}}$ within LGBCEE when prior treatments are included in the models, but stratified LTMLE otherwise.  Similar to \citep{talbot2022generalized}, the posterior probability of a given adjustment set $P(\alpha^{Y},\alpha^{Q_{1}}|\mbox{O})$ is estimated using the Metropolis-Hastings ratios calculated in \textit{step 4}. To do so, we first calculate ratios comparing the posterior probability of each explored model to a single alternative model (by multiplying together ratios comparing different models) and then divide each of these ratios by their sum. 

\textit{Step 6}. Based on \citep{raftery1993bayesian}, compute the posterior expectation $\mathbb{E}(\Delta|\mbox{O})=\sum_{\alpha^{Y},\alpha^{Q_{1}}}\hat{\Delta}^{\alpha^{Y},\alpha^{Q_{1}}}P(\alpha^{Y},\alpha^{Q_{1}}|\mbox{O})$ and the posterior variance $\mbox{Var}(\Delta|\mbox{O})=\sum_{\alpha^{Y},\alpha^{Q_{1}}}[\widehat{\mbox{Var}}(\hat{\Delta}^{\alpha^{Y},\alpha^{Q_{1}}})+(\hat{\Delta}^{\alpha^{Y},\alpha^{Q_{1}}})^{2}]{P}(\alpha^{Y},\alpha^{Q_{1}}|\mbox{O})-{\mathbb{E}}[\Delta|\mbox{O}]^{2}.$ Using a maximum likelihood approximation, the posterior distributions of the model-specific $\hat{\Delta}^{\alpha^{Y},\alpha^{Q_{1}}}$ are independent $N(\hat{\Delta}^{\alpha^{Y},\alpha^{Q_{1}}},\widehat{\mbox{Var}}(\hat{\Delta}^{\alpha^{Y},\alpha^{Q_{1}}}))$. The posterior distribution of the marginal causal effect is thus $N(\mathbb{E}(\Delta|\mbox{O}),\mbox{Var}(\Delta|\mbox{O}))$.

\section{Simulation Studies}
\subsection{Scenarios}
In this section, we describe scenarios for evaluating the performance of our algorithm, two of which are taken from \citep{schnitzer2020data} (Scenarios 1 and 3), one is an extension of Scenario 1 of \citep{talbot2022generalized} (Scenario 2), and one is an extension of a scenario in \citep{koch2018covariate} (Scenario 4). In these scenarios, we consider the treatment to be binary and the outcome to be either continuous (Scenarios 1, 2, 3 and 4) or binary (Scenarios 1B and 4B). In all scenarios, we sampled i.i.d. data $\mbox{O}=(X_0,A_0,X_1,A_1,Y)$ in sequence. The complete data generating mechanism for all scenarios is given in the Online Supplement.

\textit{Scenario 1 (Simple setting):} 
We simulated a vector $X_0=\{C_0,P_0, I_0,U_0\}$ containing four independent Gaussian covariates at time $t=0$, including an instrumental variable $I_0$ (affecting only the binary treatment choices $A_0$ and $A_1$), a variable $P_0$ directly affecting the continuous $Y$ without affecting treatment, a baseline confounding factor $C_0$ (directly affecting $A_0$, $X_1$ and $Y$), and $U_0$ which is neither affecting $A_0$ nor affecting $Y$. At time $t=1$, we simulated an instrumental variable $I_1$ (affecting treatment choice $A_1$ only), a pure predictor of outcome $P_1$ (affecting outcome only) a time-dependent confounder $C_1$ (affecting $A_1$ and $Y$) influenced by previous treatment $A_0$, and $U_1$ which is neither affecting $A_1$ nor affecting $Y$. 

\textit{Scenario 2 (Conditional instruments):} In this scenario, we simulated conditional instruments, that is, variables that are only associated with the treatment conditional on other variables. In addition we simulated two confounders that have a weak association with the treatment and outcome at each time-point.

\textit{Scenario 3 (Correlated covariates):} 
A total of $40$ baseline covariates ($X_0$) were generated from a multivariate Gaussian distribution. The covariates have variances equal to $1$ and the covariances between each pair of covariates are fixed at $0.2$, introducing moderate dependence. Among these $40$ covariates, ten were confounders, ten strongly affected treatment but less so the outcome, ten only affected the outcome, and ten affected neither the treatment nor the outcome. At time $t=1$, ten intermediate Gaussian covariates were generated conditional on the baseline covariates and treatment $A_0$, including five confounders, two were more strongly related to the treatment, and three only caused the outcome. At the end of the study, the outcome $Y$ was generated as a continuous variable conditional on treatments ($A_0, A_1$) and covariates ($X_0,X_1$).

\textit{Scenario 4 (Nonlinear outcome model):} The simulated data include five moderately correlated baseline covariates $X_0$, comprising two instrumental variables, one pure predictor of the outcome, and two confounders. At time $t=1$, additional covariates $X_1$ were generated as time‐varying confounders affected by prior treatment. The treatment assignment mechanisms at both time points depend only on linear main effects, whereas the outcome model includes quadratic and interaction terms based on both baseline and time‐varying covariates. This scenario aims to illustrate the double-robustness of LGBCEE. 

We only adapted Scenarios 1 and 4 to the binary outcome case (Scenarios 1B, 4B, respectively) due to the associated computational burden. In this case, only the equations used for generating $Y$ were changed (see Online Supplement). 

\subsection{Analysis}
First, we assessed whether the positivity assumption was met in each of the four scenarios by estimating $P(A_t=1|Z_t)$ for $t = 0,1$ with a sample size of $n = 1,000,000$. We have considered two model specifications: in the first specification, we consider ${Z}_t$, $t = 0,1$ as the set of the true direct causes of $A_t$, $t = 0,1$, and in the second specification ${Z}_t$, $t = 0,1$ as the set of the pure outcome predictors and the true confounders (target set). We then produced box plots of $\hat{P}( A_t=1|{Z}_t),t = 0,1$ for each exposure group. 

Secondly, we compared the performance of the LGBCEE and C-LTMLE methods for estimating a causal parameter $\Delta$ by generating $1,000$ datasets for each scenario. In order to allow meaningful comparisons of LGBCEE, C-LTMLE and others methods, we considered the counterfactual mean $\Delta = \mathbb{E}(Y^{11})$ because the available function implementing C-LTMLE currently only allows estimating counterfactual means. For LGBCEE, we performed $2,000$ iterations of the Markov chain Monte Carlo model composition algorithm and we set $\omega= 500\sqrt{n}$. We did not vary$\omega$ since the choice of $c \in [100, 1000]$ has little impact on GBCEE's performance \citep{talbot2022generalized}. 

For benchmark purposes, we further estimated $\Delta$ using G-computation and LTMLE and we considered two types of regression model specifications. In the first specification, we fitted the models including all potential confounders (full-GCOMP and full-LTMLE, respectively) and in the second, we only included pure outcome predictors and true confounders (target-GCOMP and target-LTMLE, respectively). 

For all methods, linear or logistic regression models were used for outcomes, and a logistic model for the treatment. Only the main covariate terms were considered, except in Scenario 3 where $A_0X_{01}$ and $A_1X_{12}$ terms were included in the outcome model for target-G, full-G, target-LTMLE, and full-LTMLE to reflect the data-generating equations. In addition, parametric likelihoods were used for calculating the BICs in LGBCEE. We implemented G-computation and LTMLE using the default options available in R's ltmle package. C-LTMLE was implemented using the code provided in \citep{schnitzer2020data}. 

For each scenario, we first calculated the true $\Delta$ using a Monte Carlo simulation where we generated a sample of $n = 1,000,000$ observations in which the counterfactual outcome $Y^{11}$ was simulated for each observation, and the true $\Delta$ was estimated by taking the mean of $Y^{11}$. The true $\Delta$s were approximately $0.286$, $0.23$, $0.6$ and $29.96$ in Scenarios 1, 2, 3 and 4, respectively. For each method and each scenario, we then calculated the bias as the difference between the mean of the estimates and the true effect, the standard deviation of the estimates (SD), the mean estimated standard error (ESE), the root mean square error (RMSE), and the proportion of simulation replicates in which the $95\%$ credible/confidence intervals included the true effect (CP). For G-computation, we estimated the variance using the non-parametric boostrap with $50$ repetitions. 
The inclusion probability of each covariate was calculated in all scenarios for LGBCEE. 

\subsection{Results}
\begin{figure}[h!]
	\centerline{\includegraphics[width=10in,width=\textwidth]{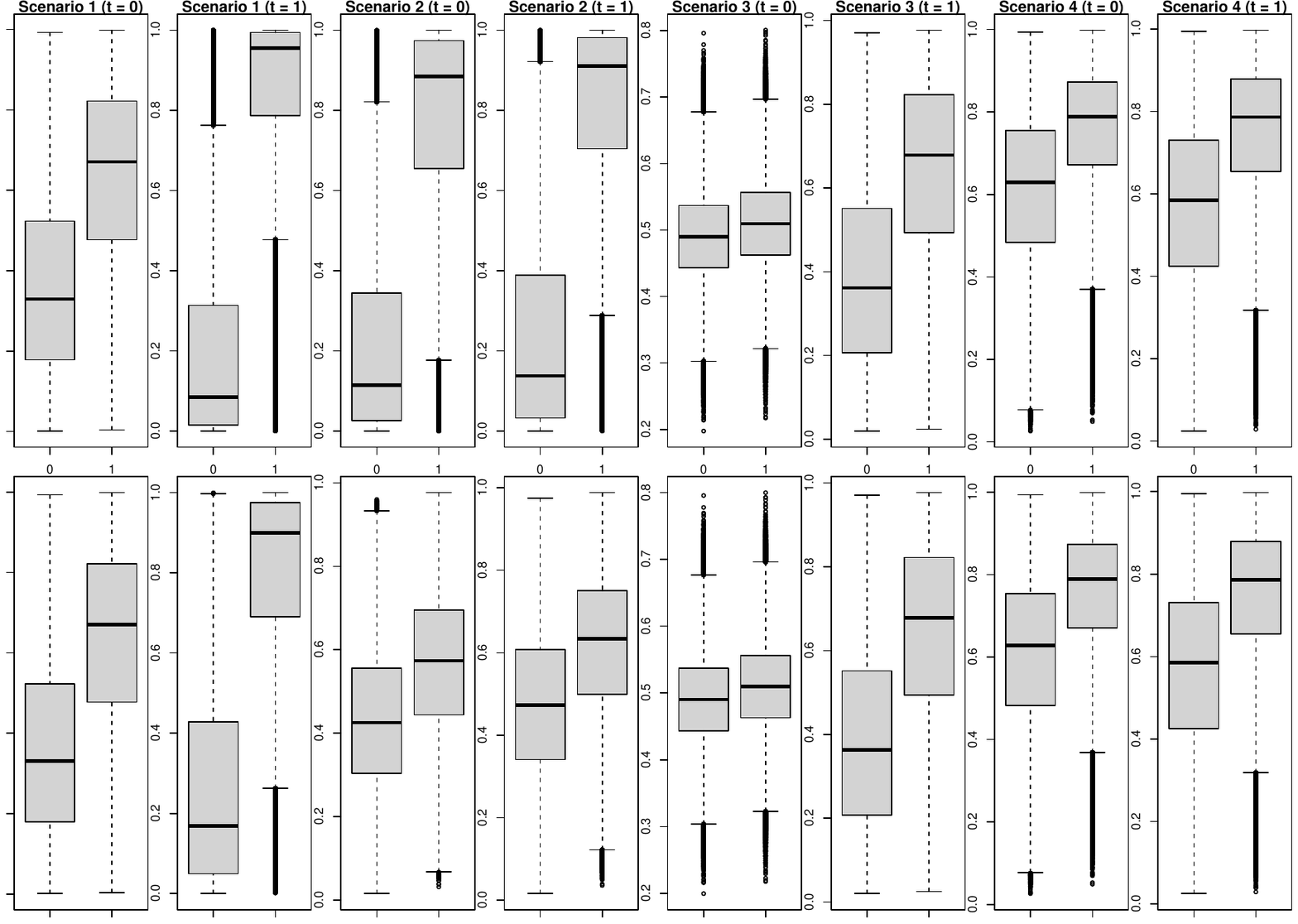}}
	\caption{Boxplots of $\hat{P}(A_t=1|{Z}_t),t = 0,1$ for each exposure group. The first row corresponds to the true exposure model by taking ${Z}_t$ as the set of the true direct causes of $A_t, t = 0, 1$ and the second row corresponds to the target exposure model by taking ${Z}_t$ as the set of the
pure outcome predictors and the true confounders.\label{fig7.1}}
\end{figure}

Figure \ref{fig7.1} shows the results of the positivity assumption verification. According to this figure, the positivity assumption would not be satisfied in Scenarios 1 and 2 if the data-generating exposure model were used. Instead considering the target adjustment set reduces positivity problems in Scenarios 2, but violations persist in Scenario 1. \\

Simulation results for Scenarios 1 to 4, are presented in Tables \ref{tab7.2} and \ref{tab7.3}. First we notice that the benchmark methods generally show negligible biases, except in Scenario 4, where Target G-COMP and Full G-COMP show higher biases. 
Secondly, Target-LTMLE and Target G-COMP generally achieve lower SD, ESE and RMSE than the Full-LTMLE and Full G-COMP methods in Scenarios 1 to 3. In Scenario 4, however, similar results are obtained for Target-LTMLE and Full-LTMLE on the one hand, and for Target G-COMP and Full G-COMP on the other hand. In this setting, the increase in SD, ESE and RMSE is particularly pronounced for Target-LTMLE and Full-LTMLE. This result showcase the sensitivity of doubly robust methods to violations of the positivity assumption.

When comparing LGBCEE and C-LTMLE, LGBCEE shows results with little or no bias. On the other hand, C-LTMLE generally shows larger biases, except for Scenario 3 where it shows a negligible bias. LGBCEE generally has lower or similar SD, ESE and RMSE than C-LTMLE. CLGBCEE shows performance in terms of bias, SD, ESE and RMSE similar to Target-LTMLE. Furthermore, LGBCEE outperforms Full-LTMLE in terms of bias, SD, ESE and RMSE.

The $95\%$ confidence interval coverage probabilities for Target-LTMLE, Full-LTMLE, Target G-COMP and Full G-COMP are generally close to $95\%$ in all scenarios.
The probability of coverage for C-TMLE was well below $95\%$ in all scenarios, except Scenario 3. This is explained by the fact that the C-TMLE method produced biased estimates. For LGBCEE, coverage was close to $95\%$ in all scenarios, even in Scenario 1 where the positivity assumption was violated. This provides empirical evidence that LGBCEE yields adequate quantification of the uncertainty associated with variable selection. 

\begin{table}
  \caption{Scenario 1 and Scenario 2 simulation study results for continuous outcome}
  \centering \scalebox{0.8}{\begin{threeparttable}
    \begin{tabular}{rrrrrrrrrrrrrr}
\cmidrule{2-13}          & \multicolumn{6}{c}{\textbf{Scenario 1}}       &       & \multicolumn{5}{c}{\textbf{Scenario 2}} &  \\
\cmidrule{2-13}          & \multicolumn{6}{c}{$n = 250$}          &       & \multicolumn{5}{c}{$n = 250$}  &  \\
\cmidrule{2-7}\cmidrule{9-13}          & \multicolumn{1}{l}{\textbf{Method}} & \multicolumn{1}{c}{\textbf{Bias}} & \multicolumn{1}{c}{\textbf{SD}} & \multicolumn{1}{c}{\textbf{ESE}} & \multicolumn{1}{c}{\textbf{RMSE}} & \multicolumn{1}{c}{\textbf{CP}} &       & \multicolumn{1}{c}{\textbf{Bias}} & \multicolumn{1}{c}{\textbf{SD}} & \multicolumn{1}{c}{\textbf{ESE}} & \multicolumn{1}{c}{\textbf{RMSE}} & \multicolumn{1}{c}{\textbf{CP}} &  \\
          & \multicolumn{1}{l}{LGBCEE} & \multicolumn{1}{c}{0.01} & \multicolumn{1}{c}{0.27} & \multicolumn{1}{c}{0.24} & \multicolumn{1}{c}{0.27} & \multicolumn{1}{c}{0.92} &       & \multicolumn{1}{c}{-0.02} & \multicolumn{1}{c}{0.23} & \multicolumn{1}{c}{0.21} & \multicolumn{1}{c}{0.22} & \multicolumn{1}{c}{0.96} &  \\
          & \multicolumn{1}{l}{Target-LTMLE} & \multicolumn{1}{c}{0.00} & \multicolumn{1}{c}{0.25} & \multicolumn{1}{c}{0.22} & \multicolumn{1}{c}{0.25} & \multicolumn{1}{c}{0.92} &       & \multicolumn{1}{c}{0.00} & \multicolumn{1}{c}{0.21} & \multicolumn{1}{c}{0.20} & \multicolumn{1}{c}{0.21} & \multicolumn{1}{c}{0.95} &  \\
          & \multicolumn{1}{l}{Full-LTMLE} & \multicolumn{1}{c}{0.00} & \multicolumn{1}{c}{0.34} & \multicolumn{1}{c}{0.26} & \multicolumn{1}{c}{0.34} & \multicolumn{1}{c}{0.92} &       & \multicolumn{1}{c}{-0.02} & \multicolumn{1}{c}{0.36} & \multicolumn{1}{c}{0.25} & \multicolumn{1}{c}{0.36} & \multicolumn{1}{c}{0.84} &  \\
          & \multicolumn{1}{l}{Target G-COMP} & \multicolumn{1}{c}{0.01} & \multicolumn{1}{c}{0.21} & \multicolumn{1}{c}{0.20} & \multicolumn{1}{c}{0.21} & \multicolumn{1}{c}{0.93} &       & \multicolumn{1}{c}{-0.01} & \multicolumn{1}{c}{0.19} & \multicolumn{1}{c}{0.18} & \multicolumn{1}{c}{0.19} & \multicolumn{1}{c}{0.95} &  \\
          & \multicolumn{1}{l}{Full G-COMP} & \multicolumn{1}{c}{0.01} & \multicolumn{1}{c}{0.22} & \multicolumn{1}{c}{0.21} & \multicolumn{1}{c}{0.22} & \multicolumn{1}{c}{0.93} &       & \multicolumn{1}{c}{-0.01} & \multicolumn{1}{c}{0.21} & \multicolumn{1}{c}{0.21} & \multicolumn{1}{c}{0.21} & \multicolumn{1}{c}{0.94} &  \\
          & \multicolumn{1}{l}{C-LTMLE} & \multicolumn{1}{c}{0.56} & \multicolumn{1}{c}{0.28} & \multicolumn{1}{c}{0.56} & \multicolumn{1}{c}{0.63} & \multicolumn{1}{c}{0.45} &       & \multicolumn{1}{c}{-0.02} & \multicolumn{1}{c}{0.22} & \multicolumn{1}{c}{0.22} & \multicolumn{1}{c}{0.22} & \multicolumn{1}{c}{0.29} &  \\
\cmidrule{2-7}\cmidrule{9-13}          & \multicolumn{6}{c}{$n = 500$}          &       & \multicolumn{5}{c}{$n = 500$}  &  \\
\cmidrule{2-7}\cmidrule{9-13}          & \multicolumn{1}{l}{\textbf{Method}} & \multicolumn{1}{c}{\textbf{Bias}} & \multicolumn{1}{c}{\textbf{SD}} & \multicolumn{1}{c}{\textbf{ESE}} & \multicolumn{1}{c}{\textbf{RMSE}} & \multicolumn{1}{c}{\textbf{CP}} &       & \multicolumn{1}{c}{\textbf{Bias}} & \multicolumn{1}{c}{\textbf{SD}} & \multicolumn{1}{c}{\textbf{ESE}} & \multicolumn{1}{c}{\textbf{RMSE}} & \multicolumn{1}{c}{\textbf{CP}} &  \\
          & \multicolumn{1}{l}{LGBCEE} & \multicolumn{1}{c}{0.01} & \multicolumn{1}{c}{0.19} & \multicolumn{1}{c}{0.18} & \multicolumn{1}{c}{0.19} & \multicolumn{1}{c}{0.94} &       & \multicolumn{1}{c}{-0.01} & \multicolumn{1}{c}{0.15} & \multicolumn{1}{c}{0.14} & \multicolumn{1}{c}{0.14} & \multicolumn{1}{c}{0.97} &  \\
          & \multicolumn{1}{l}{Target-LTMLE} & \multicolumn{1}{c}{0.01} & \multicolumn{1}{c}{0.18} & \multicolumn{1}{c}{0.16} & \multicolumn{1}{c}{0.18} & \multicolumn{1}{c}{0.93} &       & \multicolumn{1}{c}{-0.01} & \multicolumn{1}{c}{0.14} & \multicolumn{1}{c}{0.14} & \multicolumn{1}{c}{0.14} & \multicolumn{1}{c}{0.95} &  \\
          & \multicolumn{1}{l}{Full-LTMLE} & \multicolumn{1}{c}{0.01} & \multicolumn{1}{c}{0.25} & \multicolumn{1}{c}{0.20} & \multicolumn{1}{c}{0.25} & \multicolumn{1}{c}{0.93} &       & \multicolumn{1}{c}{-0.01} & \multicolumn{1}{c}{0.31} & \multicolumn{1}{c}{0.22} & \multicolumn{1}{c}{0.31} & \multicolumn{1}{c}{0.90} &  \\
          & \multicolumn{1}{l}{Target G-COMP} & \multicolumn{1}{c}{0.02} & \multicolumn{1}{c}{0.15} & \multicolumn{1}{c}{0.14} & \multicolumn{1}{c}{0.15} & \multicolumn{1}{c}{0.93} &       & \multicolumn{1}{c}{-0.02} & \multicolumn{1}{c}{0.13} & \multicolumn{1}{c}{0.13} & \multicolumn{1}{c}{0.13} & \multicolumn{1}{c}{0.94} &  \\
          & \multicolumn{1}{l}{Full G-COMP} & \multicolumn{1}{c}{0.01} & \multicolumn{1}{c}{0.16} & \multicolumn{1}{c}{0.15} & \multicolumn{1}{c}{0.16} & \multicolumn{1}{c}{0.93} &       & \multicolumn{1}{c}{-0.02} & \multicolumn{1}{c}{0.14} & \multicolumn{1}{c}{0.15} & \multicolumn{1}{c}{0.15} & \multicolumn{1}{c}{0.95} &  \\
          & \multicolumn{1}{l}{C-LTMLE} & \multicolumn{1}{c}{0.56} & \multicolumn{1}{c}{0.22} & \multicolumn{1}{c}{0.35} & \multicolumn{1}{c}{0.60} & \multicolumn{1}{c}{0.28} &       & \multicolumn{1}{c}{0.17} & \multicolumn{1}{c}{0.14} & \multicolumn{1}{c}{0.13} & \multicolumn{1}{c}{0.22} & \multicolumn{1}{c}{0.37} &  \\
\cmidrule{2-7}\cmidrule{9-13}          & \multicolumn{6}{c}{$n = 1000$}         &       & \multicolumn{5}{c}{$n = 1000$} &  \\
\cmidrule{2-7}\cmidrule{9-13}          & \multicolumn{1}{l}{\textbf{Method}} & \multicolumn{1}{c}{\textbf{Bias}} & \multicolumn{1}{c}{\textbf{SD}} & \multicolumn{1}{c}{\textbf{ESE}} & \multicolumn{1}{c}{\textbf{RMSE}} & \multicolumn{1}{c}{\textbf{CP}} &       & \multicolumn{1}{c}{\textbf{Bias}} & \multicolumn{1}{c}{\textbf{SD}} & \multicolumn{1}{c}{\textbf{ESE}} & \multicolumn{1}{c}{\textbf{RMSE}} & \multicolumn{1}{c}{\textbf{CP}} &  \\
          & \multicolumn{1}{l}{LGBCEE} & \multicolumn{1}{c}{0.01} & \multicolumn{1}{c}{0.13} & \multicolumn{1}{c}{0.12} & \multicolumn{1}{c}{0.13} & \multicolumn{1}{c}{0.92} &       & \multicolumn{1}{c}{0.00} & \multicolumn{1}{c}{0.11} & \multicolumn{1}{c}{0.10} & \multicolumn{1}{c}{0.10} & \multicolumn{1}{c}{0.96} &  \\
          & \multicolumn{1}{l}{Target-LTMLE} & \multicolumn{1}{c}{0.02} & \multicolumn{1}{c}{0.12} & \multicolumn{1}{c}{0.11} & \multicolumn{1}{c}{0.12} & \multicolumn{1}{c}{0.92} &       & \multicolumn{1}{c}{0.00} & \multicolumn{1}{c}{0.10} & \multicolumn{1}{c}{0.10} & \multicolumn{1}{c}{0.10} & \multicolumn{1}{c}{0.95} &  \\
          & \multicolumn{1}{l}{Full-LTMLE} & \multicolumn{1}{c}{0.02} & \multicolumn{1}{c}{0.21} & \multicolumn{1}{c}{0.16} & \multicolumn{1}{c}{0.21} & \multicolumn{1}{c}{0.92} &       & \multicolumn{1}{c}{0.00} & \multicolumn{1}{c}{0.23} & \multicolumn{1}{c}{0.18} & \multicolumn{1}{c}{0.23} & \multicolumn{1}{c}{0.91} &  \\
          & \multicolumn{1}{l}{Target G-COMP} & \multicolumn{1}{c}{0.02} & \multicolumn{1}{c}{0.11} & \multicolumn{1}{c}{0.10} & \multicolumn{1}{c}{0.11} & \multicolumn{1}{c}{0.94} &       & \multicolumn{1}{c}{-0.01} & \multicolumn{1}{c}{0.09} & \multicolumn{1}{c}{0.09} & \multicolumn{1}{c}{0.09} & \multicolumn{1}{c}{0.95} &  \\
          & \multicolumn{1}{l}{Full G-COMP} & \multicolumn{1}{c}{0.02} & \multicolumn{1}{c}{0.11} & \multicolumn{1}{c}{0.10} & \multicolumn{1}{c}{0.11} & \multicolumn{1}{c}{0.93} &       & \multicolumn{1}{c}{-0.01} & \multicolumn{1}{c}{0.10} & \multicolumn{1}{c}{0.10} & \multicolumn{1}{c}{0.10} & \multicolumn{1}{c}{0.95} &  \\
          & \multicolumn{1}{l}{C-LTMLE} & \multicolumn{1}{c}{0.56} & \multicolumn{1}{c}{0.16} & \multicolumn{1}{c}{0.28} & \multicolumn{1}{c}{0.58} & \multicolumn{1}{c}{0.23} &       & \multicolumn{1}{c}{0.18} & \multicolumn{1}{c}{0.10} & \multicolumn{1}{c}{0.10} & \multicolumn{1}{c}{0.20} & \multicolumn{1}{c}{0.51} &  \\
\cmidrule{2-13} 
    \end{tabular}%
    \begin{tablenotes}
 \item Scenario 1 is the simple scenario simulation study, Scenario 2 is the conditional instruments case.
		\end{tablenotes}
\end{threeparttable}}
 \label{tab7.2}%
\end{table}%

\begin{table}
\caption{Scenario 3 and Scenario 4 simulation study results for continuous outcome}
\centering \scalebox{0.8}{\begin{threeparttable}
    \begin{tabular}{rrrrrrrrrrrrrr}
\cmidrule{2-13}          & \multicolumn{6}{c}{\textbf{Scenario 3}}       &       & \multicolumn{5}{c}{\textbf{Scenario 4}} &  \\
\cmidrule{2-13}          & \multicolumn{6}{c}{$n = 250$}          &       & \multicolumn{5}{c}{$n = 250$}  &  \\
\cmidrule{2-7}\cmidrule{9-13}          & \multicolumn{1}{l}{\textbf{Method}} & \multicolumn{1}{c}{\textbf{Bias}} & \multicolumn{1}{c}{\textbf{SD}} & \multicolumn{1}{c}{\textbf{ESE}} & \multicolumn{1}{c}{\textbf{RMSE}} & \multicolumn{1}{c}{\textbf{CP}} &       & \multicolumn{1}{c}{\textbf{Bias}} & \multicolumn{1}{c}{\textbf{SD}} & \multicolumn{1}{c}{\textbf{ESE}} & \multicolumn{1}{c}{\textbf{RMSE}} & \multicolumn{1}{c}{\textbf{CP}} &  \\
          & \multicolumn{1}{l}{LGBCEE} & \multicolumn{1}{c}{0.01} & \multicolumn{1}{c}{0.18} & \multicolumn{1}{c}{0.18} & \multicolumn{1}{c}{0.18} & \multicolumn{1}{c}{0.95} &       & \multicolumn{1}{c}{-0.08} & \multicolumn{1}{c}{2.41} & \multicolumn{1}{c}{2.37} & \multicolumn{1}{c}{2.41} & \multicolumn{1}{c}{0.95} &  \\
          & \multicolumn{1}{l}{Target-LTMLE} & \multicolumn{1}{c}{0.00} & \multicolumn{1}{c}{0.21} & \multicolumn{1}{c}{0.18} & \multicolumn{1}{c}{0.21} & \multicolumn{1}{c}{0.92} &       & \multicolumn{1}{c}{-0.12} & \multicolumn{1}{c}{2.53} & \multicolumn{1}{c}{2.38} & \multicolumn{1}{c}{2.53} & \multicolumn{1}{c}{0.94} &  \\
          & \multicolumn{1}{l}{Full-LTMLE} & \multicolumn{1}{c}{0.01} & \multicolumn{1}{c}{0.23} & \multicolumn{1}{c}{0.20} & \multicolumn{1}{c}{0.23} & \multicolumn{1}{c}{0.91} &       & \multicolumn{1}{c}{-0.12} & \multicolumn{1}{c}{2.53} & \multicolumn{1}{c}{2.38} & \multicolumn{1}{c}{2.53} & \multicolumn{1}{c}{0.94} &  \\
          & \multicolumn{1}{l}{Target G-COMP} & \multicolumn{1}{c}{-0.03} & \multicolumn{1}{c}{0.15} & \multicolumn{1}{c}{0.15} & \multicolumn{1}{c}{0.16} & \multicolumn{1}{c}{0.94} &       & \multicolumn{1}{c}{-0.31} & \multicolumn{1}{c}{2.21} & \multicolumn{1}{c}{2.11} & \multicolumn{1}{c}{2.23} & \multicolumn{1}{c}{0.93} &  \\
          & \multicolumn{1}{l}{Full G-COMP} & \multicolumn{1}{c}{-0.02} & \multicolumn{1}{c}{0.16} & \multicolumn{1}{c}{0.16} & \multicolumn{1}{c}{0.16} & \multicolumn{1}{c}{0.93} &       & \multicolumn{1}{c}{-0.31} & \multicolumn{1}{c}{2.21} & \multicolumn{1}{c}{2.11} & \multicolumn{1}{c}{2.23} & \multicolumn{1}{c}{0.93} &  \\
          & \multicolumn{1}{l}{C-LTMLE} & \multicolumn{1}{c}{-0.02} & \multicolumn{1}{c}{0.17} & \multicolumn{1}{c}{0.17} & \multicolumn{1}{c}{0.17} & \multicolumn{1}{c}{0.94} &       & \multicolumn{1}{c}{4.04} & \multicolumn{1}{c}{3.17} & \multicolumn{1}{c}{2.95} & \multicolumn{1}{c}{5.13} & \multicolumn{1}{c}{0.76} &  \\
\cmidrule{2-7}\cmidrule{9-13}          & \multicolumn{6}{c}{$n = 500$}          &       & \multicolumn{5}{c}{$n = 500$}  &  \\
\cmidrule{2-7}\cmidrule{9-13}          & \multicolumn{1}{l}{\textbf{Method}} & \multicolumn{1}{c}{\textbf{Bias}} & \multicolumn{1}{c}{\textbf{SD}} & \multicolumn{1}{c}{\textbf{ESE}} & \multicolumn{1}{c}{\textbf{RMSE}} & \multicolumn{1}{c}{\textbf{CP}} &       & \multicolumn{1}{c}{\textbf{Bias}} & \multicolumn{1}{c}{\textbf{SD}} & \multicolumn{1}{c}{\textbf{ESE}} & \multicolumn{1}{c}{\textbf{RMSE}} & \multicolumn{1}{c}{\textbf{CP}} &  \\
          & \multicolumn{1}{l}{LGBCEE} & \multicolumn{1}{c}{-0.01} & \multicolumn{1}{c}{0.13} & \multicolumn{1}{c}{0.13} & \multicolumn{1}{c}{0.13} & \multicolumn{1}{c}{0.93} &       & \multicolumn{1}{c}{0.03} & \multicolumn{1}{c}{2.04} & \multicolumn{1}{c}{1.76} & \multicolumn{1}{c}{2.04} & \multicolumn{1}{c}{0.96} &  \\
          & \multicolumn{1}{l}{Target-LTMLE} & \multicolumn{1}{c}{0.00} & \multicolumn{1}{c}{0.14} & \multicolumn{1}{c}{0.13} & \multicolumn{1}{c}{0.14} & \multicolumn{1}{c}{0.94} &       & \multicolumn{1}{c}{-0.01} & \multicolumn{1}{c}{2.03} & \multicolumn{1}{c}{1.75} & \multicolumn{1}{c}{2.03} & \multicolumn{1}{c}{0.95} &  \\
          & \multicolumn{1}{l}{Full-LTMLE} & \multicolumn{1}{c}{-0.01} & \multicolumn{1}{c}{0.15} & \multicolumn{1}{c}{0.14} & \multicolumn{1}{c}{0.15} & \multicolumn{1}{c}{0.93} &       & \multicolumn{1}{c}{-0.01} & \multicolumn{1}{c}{2.03} & \multicolumn{1}{c}{1.75} & \multicolumn{1}{c}{2.03} & \multicolumn{1}{c}{0.95} &  \\
          & \multicolumn{1}{l}{Target G-COMP} & \multicolumn{1}{c}{-0.03} & \multicolumn{1}{c}{0.11} & \multicolumn{1}{c}{0.11} & \multicolumn{1}{c}{0.11} & \multicolumn{1}{c}{0.93} &       & \multicolumn{1}{c}{-0.27} & \multicolumn{1}{c}{1.46} & \multicolumn{1}{c}{1.50} & \multicolumn{1}{c}{1.48} & \multicolumn{1}{c}{0.95} &  \\
          & \multicolumn{1}{l}{Full G-COMP} & \multicolumn{1}{c}{-0.03} & \multicolumn{1}{c}{0.11} & \multicolumn{1}{c}{0.11} & \multicolumn{1}{c}{0.12} & \multicolumn{1}{c}{0.93} &       & \multicolumn{1}{c}{-0.27} & \multicolumn{1}{c}{1.46} & \multicolumn{1}{c}{1.50} & \multicolumn{1}{c}{1.48} & \multicolumn{1}{c}{0.95} &  \\
          & \multicolumn{1}{l}{C-LTMLE} & \multicolumn{1}{c}{-0.03} & \multicolumn{1}{c}{0.13} & \multicolumn{1}{c}{0.12} & \multicolumn{1}{c}{0.13} & \multicolumn{1}{c}{0.92} &       & \multicolumn{1}{c}{4.39} & \multicolumn{1}{c}{2.25} & \multicolumn{1}{c}{2.19} & \multicolumn{1}{c}{4.93} & \multicolumn{1}{c}{0.50} &  \\
\cmidrule{2-7}\cmidrule{9-13}          & \multicolumn{6}{c}{$n = 1000$}         &       & \multicolumn{5}{c}{$n = 1000$} &  \\
\cmidrule{2-7}\cmidrule{9-13}          & \multicolumn{1}{l}{\textbf{Method}} & \multicolumn{1}{c}{\textbf{Bias}} & \multicolumn{1}{c}{\textbf{SD}} & \multicolumn{1}{c}{\textbf{ESE}} & \multicolumn{1}{c}{\textbf{RMSE}} & \multicolumn{1}{c}{\textbf{CP}} &       & \multicolumn{1}{c}{\textbf{Bias}} & \multicolumn{1}{c}{\textbf{SD}} & \multicolumn{1}{c}{\textbf{ESE}} & \multicolumn{1}{c}{\textbf{RMSE}} & \multicolumn{1}{c}{\textbf{CP}} &  \\
          & \multicolumn{1}{l}{LGBCEE} & \multicolumn{1}{c}{0.00} & \multicolumn{1}{c}{0.09} & \multicolumn{1}{c}{0.09} & \multicolumn{1}{c}{0.09} & \multicolumn{1}{c}{0.95} &       & \multicolumn{1}{c}{0.03} & \multicolumn{1}{c}{1.31} & \multicolumn{1}{c}{1.27} & \multicolumn{1}{c}{1.30} & \multicolumn{1}{c}{0.96} &  \\
          & \multicolumn{1}{l}{Target-LTMLE} & \multicolumn{1}{c}{0.00} & \multicolumn{1}{c}{0.09} & \multicolumn{1}{c}{0.09} & \multicolumn{1}{c}{0.09} & \multicolumn{1}{c}{0.95} &       & \multicolumn{1}{c}{0.02} & \multicolumn{1}{c}{1.31} & \multicolumn{1}{c}{1.27} & \multicolumn{1}{c}{1.31} & \multicolumn{1}{c}{0.96} &  \\
          & \multicolumn{1}{l}{Full-LTMLE} & \multicolumn{1}{c}{-0.01} & \multicolumn{1}{c}{0.10} & \multicolumn{1}{c}{0.10} & \multicolumn{1}{c}{0.10} & \multicolumn{1}{c}{0.94} &       & \multicolumn{1}{c}{0.02} & \multicolumn{1}{c}{1.31} & \multicolumn{1}{c}{1.27} & \multicolumn{1}{c}{1.31} & \multicolumn{1}{c}{0.96} &  \\
          & \multicolumn{1}{l}{Target G-COMP} & \multicolumn{1}{c}{-0.02} & \multicolumn{1}{c}{0.08} & \multicolumn{1}{c}{0.08} & \multicolumn{1}{c}{0.08} & \multicolumn{1}{c}{0.94} &       & \multicolumn{1}{c}{-0.27} & \multicolumn{1}{c}{1.05} & \multicolumn{1}{c}{1.06} & \multicolumn{1}{c}{1.09} & \multicolumn{1}{c}{0.94} &  \\
          & \multicolumn{1}{l}{Full G-COMP} & \multicolumn{1}{c}{-0.03} & \multicolumn{1}{c}{0.08} & \multicolumn{1}{c}{0.08} & \multicolumn{1}{c}{0.08} & \multicolumn{1}{c}{0.93} &       & \multicolumn{1}{c}{-0.27} & \multicolumn{1}{c}{1.05} & \multicolumn{1}{c}{1.06} & \multicolumn{1}{c}{1.09} & \multicolumn{1}{c}{0.94} &  \\
          & \multicolumn{1}{l}{C-LTMLE} & \multicolumn{1}{c}{-0.03} & \multicolumn{1}{c}{0.09} & \multicolumn{1}{c}{0.09} & \multicolumn{1}{c}{0.09} & \multicolumn{1}{c}{0.96} &       & \multicolumn{1}{c}{4.50} & \multicolumn{1}{c}{1.60} & \multicolumn{1}{c}{1.59} & \multicolumn{1}{c}{4.77} & \multicolumn{1}{c}{0.16} &  \\
\cmidrule{2-13}     
    \end{tabular}%
    \begin{tablenotes}
 \item Scenario 3 is the correlated covariates, Scenario 4 is the continuous outcome with nonlinear outcome model.
\end{tablenotes}
\end{threeparttable}}
  \label{tab7.3}%
\end{table}%

The Online Supplement shows graphs of covariate inclusion probabilities. These figures reveal that LGBCEE has high probabilities of including both true confounders and outcome risk factors, and lower probabilities of including the other variables. In general, the variables selected are close to those aimed by the target for larger $n$, but less so in Scenarios 3 and 4, particularly for a binary outcome.  Analyses for a binary outcome and exposure (Scenario 1B, Scenario 4B), comparing different methods, are also shown in the Online Supplement. In Scenario 1B, all methods had low bias, SD, ESE and RMSE. The methods also showed good coverage, with the exception of C-LTMLE. Surprisingly, the results of Scenario 4B are not similar to those of Scenario 4. Specifically, all methods produced low bias, SD, ESE, RMSE and their 
$95\%$ confidence interval coverage probabilities are generally close to $95\%$.

\section{Real-data illustration}


To illustrate the use of LGBCEE, we estimated the effect of statin regimens for preventing a first cardiovascular disease (CVD) or death risk in men aged 66 or over. We built a retrospective cohort using data from the Institut de la statistique du Québec. To do this, we merged five medico-administrative databases from Quebec, Canada, using a unique anonymized personal identifier. These databases include the health insurance registry, the pharmaceutical services database, the medical claims database, the hospitalization database and the death registry. The cohort includes Quebec residents aged 66 or over on April 1, 2013, who have benefited continuously from Quebec's public drug insurance plan during the previous year. To ensure that only new primary prevention statin users were included, we excluded individuals who had obtained a statin prescription between April 1, 2012 and March 31, 2013, as well as those with a history of cardiovascular disease between April 1, 2008 and March 31, 2013. To limit confounding by indication bias, the analysis focused on people who had received at least one statin prescription between April 2012 and March 2017 \citep{mesidor2024effect}. Statin exposure and potential confounding variables were tracked and updated monthly up to 12 months after first dispensation. 

First, we created a 12-month data set containing statin exposure, potential covariates and our events of interest. In this case, a person was considered exposed during a given month if he had a statin prescription covering at least one day of that month, taking into account a $50\%$ equivalent grace period. Event follow-up began immediately after the 12-month exposure period and continued until the first of the following: occurrence of a cardiovascular event, death, termination of public drug plan membership, admission to a long-term care facility or 12 months after the end of exposure follow-up. We considered a binary outcome $Y$ measured after the exposure follow-up that takes the value $Y = 1$ if the individual experienced a cardiovascular event or died of any cause during the 1-year outcome follow-up period and $0$ otherwise. Individuals who experienced any of the following during the exposure follow-up were excluded: a cardiovascular event, death or discontinuation of public drug plan membership. We took into account the following potential confounders: age, prevalent diabetes, hypertension, chronic kidney disease, use of aspirin and other antiplatelet agents, treatment of hypertension, number of prescriptions filled in the year of statin initiation and number of hospital days in the year of statin initiation. 

We then grouped the data into four periods ($t = 0, 1, 2, 3$) of 3 months. 
Subjects were considered exposed during a given period if they were exposed to statins ($A_t = 1$) at least one day during that period, and unexposed otherwise ($A_t = 0$). 
For each of the continuous confounding factors contained in the data, we created a new corresponding variable by averaging the values of the three months of the period. For binary variables, we created a new variable equal to $1$ if a subject expressed this variable at least once during a given period and $0$ otherwise. The effect of statins was estimated using PLTMLE, adjusting for all potential confounders, and LGBCEE. 
Since all participants were statin users at the beginning of the study, we compared persistent usage ($a_t = 1$ for $t = 1, 2, 3$) to early discontinuation ($a_t = 0$ for $t = 1, 2, 3$) using the additive MSM $P(Y^{\bar{a}_3}=1) = \beta_0 + \beta_1 a_1 + \beta_2 a_2 + \beta_3 a_3$.

The Online Supplement presents a comparison of the baseline characteristics of the population and correlation between variables respectively. The study sample included $14,020$ individuals, with a mean age of $71.59$ years ($SD = 5.37$). Among the $14,020$ men included, $1,779$ participants had a censoring event during treatment follow-up and were excluded from the analysis. With regard to comorbidities and initial treatments, $0.7\%$ of participants had chronic kidney disease at baseline, while $4.4\%$ were taking aspirin. 

The fully adjusted PLTMLE estimated a difference in CVD risk of $-3.9\%$ ($95\% CI: -8.3\%, 0.5\%$) between persistent statin users and early discontinuers. LGBCEE with $\omega = 500\sqrt{n}$ and $20,000$ iterations gave an estimate of $-3.1\%$ ($95\% CI: -5.8\%, -0.4\%$). This approach considerably reduced the width of the confidence/credible interval ($69\%$) compared to the fully adjusted estimate. The results of the variable selection are presented in Table~\ref{tab:74}. 
Interestingly, LGBCEE avoided simultaneously selecting variables that were highly correlated. For example, at time $t = 3$, LGBCEE selected prevalent diabetes at time $t = 2$, but excluded prevalent diabetes at time $t = 3$ and prevalent diabetes prevalent at time $t = 0$. 

\renewcommand{\arraystretch}{1.25}
\begin{table}
\caption{Probability of selection of covariates at each time-point}
  \centering \scalebox{0.7}{\begin{threeparttable}
  \begin{tabular}{rrrrrr}
\cmidrule{2-5}          & \multicolumn{1}{l}{\multirow{2}[4]{*}{\textbf{Covariates}}} & \multicolumn{3}{p{9.835em}}{\centering \textbf{Time-point}} &  \\
\cmidrule{3-5}          &       & \multicolumn{1}{p{3.335em}}{\centering $t = 1$} & \multicolumn{1}{p{2.835em}}{\centering $t = 2$} & \multicolumn{1}{p{3.665em}}{\centering $t = 3$} &  \\
\cmidrule{2-5}          & \multicolumn{1}{p{30.335em}}{Age} & \multicolumn{1}{c}{100\%} & \multicolumn{1}{c}{100\%} & \multicolumn{1}{c}{100\%} &  \\
          & \multicolumn{1}{l}{Chronic kidney disease ($t = 0$)} & \multicolumn{1}{c}{0\%} & \multicolumn{1}{c}{1\%} & \multicolumn{1}{c}{0\%} &  \\
          & \multicolumn{1}{l}{Use of aspirin and other antiplatelet agents ($t = 0$)} & \multicolumn{1}{c}{0\%} & \multicolumn{1}{c}{0\%} & \multicolumn{1}{c}{0\%} &  \\
          & \multicolumn{1}{l}{Blood pressure treatment at time ($t = 0$)} & \multicolumn{1}{c}{0\%} & \multicolumn{1}{c}{0\%} & \multicolumn{1}{c}{0\%} &  \\
          & \multicolumn{1}{l}{Number of filled prescriptions in the year preceding time ($t = 0$)} & \multicolumn{1}{c}{0\%} & \multicolumn{1}{c}{0\%} & \multicolumn{1}{c}{0\%} &  \\
          & \multicolumn{1}{l}{Number of days of hospitalization in the year preceding time ($t = 0$)} & \multicolumn{1}{c}{91\%} & \multicolumn{1}{c}{5\%} & \multicolumn{1}{c}{9\%} &  \\
          & \multicolumn{1}{l}{Hypertension ($t = 0$)} & \multicolumn{1}{c}{0\%} & \multicolumn{1}{c}{0\%} & \multicolumn{1}{c}{6\%} &  \\
          & \multicolumn{1}{l}{Prevalent diabetes ($t = 0$)} & \multicolumn{1}{c}{84\%} & \multicolumn{1}{c}{3\%} & \multicolumn{1}{c}{0\%} &  \\
          & \multicolumn{1}{l}{Chronic kidney disease ($t = 1$)} & \multicolumn{1}{c}{19\%} & \multicolumn{1}{c}{0\%} & \multicolumn{1}{c}{0\%} &  \\
          & \multicolumn{1}{l}{Use of aspirin and other antiplatelet agents ($t = 1$)} & \multicolumn{1}{c}{0\%} & \multicolumn{1}{c}{0\%} & \multicolumn{1}{c}{0\%} &  \\
          & \multicolumn{1}{l}{Blood pressure treatment ($t = 1$)} & \multicolumn{1}{c}{0\%} & \multicolumn{1}{c}{4\%} & \multicolumn{1}{c}{0\%} &  \\
          & \multicolumn{1}{l}{Number of filled prescriptions in the year preceding time ($t = 1$)} & \multicolumn{1}{c}{100\%} & \multicolumn{1}{c}{13\%} & \multicolumn{1}{c}{0\%} &  \\
          & \multicolumn{1}{l}{Number of days of hospitalization in the year preceding time ($t = 1$)} & \multicolumn{1}{c}{100\%} & \multicolumn{1}{c}{64\%} & \multicolumn{1}{c}{0\%} &  \\
          & \multicolumn{1}{l}{Hypertension ($t = 1$)} & \multicolumn{1}{c}{0\%} & \multicolumn{1}{c}{0\%} & \multicolumn{1}{c}{2\%} &  \\
          & \multicolumn{1}{l}{Prevalent diabetes ($t = 1$)} & \multicolumn{1}{c}{11\%} & \multicolumn{1}{c}{3\%} & \multicolumn{1}{c}{0\%} &  \\
          & \multicolumn{1}{p{30.335em}}{$A_1$} &       & \multicolumn{1}{c}{0\%} & \multicolumn{1}{c}{100\%} &  \\
          & \multicolumn{1}{l}{Chronic kidney disease ($t = 2$)} &       & \multicolumn{1}{c}{4\%} & \multicolumn{1}{c}{0\%} &  \\
          & \multicolumn{1}{l}{Use of aspirin and other antiplatelet agents ($t = 2$)} &       & \multicolumn{1}{c}{0\%} & \multicolumn{1}{c}{0\%} &  \\
          & \multicolumn{1}{l}{Blood pressure treatment ($t = 2$)} &       & \multicolumn{1}{c}{6\%} & \multicolumn{1}{c}{34\%} &  \\
          & \multicolumn{1}{l}{Number of filled prescriptions in the year preceding time ($t = 2$)} &       & \multicolumn{1}{c}{100\%} & \multicolumn{1}{c}{0\%} &  \\
          & \multicolumn{1}{l}{Number of days of hospitalization in the year preceding time ($t = 2$)} &       & \multicolumn{1}{c}{100\%} & \multicolumn{1}{c}{0\%} &  \\
          & \multicolumn{1}{l}{Hypertension ($t = 2$)} &       & \multicolumn{1}{c}{0\%} & \multicolumn{1}{c}{0\%} &  \\
          & \multicolumn{1}{l}{Prevalent diabetes ($t = 2$)} &       & \multicolumn{1}{c}{89\%} & \multicolumn{1}{c}{100\%} &  \\
          & \multicolumn{1}{p{30.335em}}{$A_2$} &       &       & \multicolumn{1}{c}{6\%} &  \\
          & \multicolumn{1}{l}{Chronic kidney disease ($t = 3$)} &       &       & \multicolumn{1}{c}{8\%} &  \\
          & \multicolumn{1}{l}{Use of aspirin and other antiplatelet agents ($t = 3$)} &       &       & \multicolumn{1}{c}{0\%} &  \\
          & \multicolumn{1}{l}{Blood pressure treatment ($t = 3$)} &       &       & \multicolumn{1}{c}{93\%} &  \\
          & \multicolumn{1}{l}{Number of filled prescriptions in the year of statin initiation ($t = 3$)} &       &       & \multicolumn{1}{c}{100\%} &  \\
          & \multicolumn{1}{l}{Number of days of hospitalization in the year of statin initiation ($t = 3$)} &       &       & \multicolumn{1}{c}{100\%} &  \\
          & \multicolumn{1}{l}{Hypertension ($t = 3$)} &       &       & \multicolumn{1}{c}{0\%} &  \\
          & \multicolumn{1}{l}{Prevalent diabetes ($t = 3$)} &       &       & \multicolumn{1}{c}{0\%} &  \\
\cmidrule{2-5} 
    \end{tabular}
\begin{tablenotes}   
\item Variables $A_1$ and $A_2$ represent statin use at time $t=1$ and $t=2$ respectively.
\end{tablenotes}
\end{threeparttable}} \label{tab:74}
\end{table}

\section{Discussion}
In this study, we developed a longitudinal extension to the Generalized Bayesian Causal Effect estimation algorithm, which we call LGBCEE, to estimate the causal effect of a time-varying binary treatment on a binary or continuous outcome. The proposed method is based on a Bayesian model average of LTMLE estimates. To the best of our knowledge, this constitutes the first proposal of a Bayesian variable selection approach for longitudinal causal effect estimation. 
The variable selection objective of LGBCEE is geared toward the selection of sufficient adjustment sets with reduced variance, as established in previous theoretical work in a setting with known causal graphs \citep{rotnitzky2020efficient, adenyo2024longitudinal}. However, it operationalizes these principles in a setting where a causal graph cannot be constructed based on domain knowledge alone.

Through a series of simulations, we evaluated the performance of LGBCEE against an existing model selection method, C-LTMLE, and other doubly robust methods. Our results show that LGBCEE offers improved performance as compared to Full-LTMLE and even comparable performance to the Target-LTMLE method in terms of bias, SD, ESE and RMSE in the majority of scenarios studied. In most scenarios, LGBCEE also outperformed C-LTMLE in terms of bias, while performing similarly in terms of SD. These results highlight the ability of LGBCEE to perform well without requiring a priori knowledge of the target model, which is often unrealistic in practice. 
A major strength of LGBCEE lies in its ability to produce adequate credible intervals, even with relatively small sample sizes or when the positivity assumption is violated. This property is essential for application in epidemiology and clinical research, where longitudinal studies often involve modest sample sizes. 

We applied LGBCEE to a longitudinal dataset investigating the effect of statin use regimens on CVD risk in older adults. LGBCEE produced an effect estimate comparable to that obtained by a fully adjusted doubly robust estimator, but with a significantly narrower credible/confidence interval, suggesting a significant improvement in the precision of the results due to the correct selection of confounders. 

Despite its advantages, LGBCEE has certain limitations. Firstly, the LGBCEE algorithm performs a Bayesian averaging of the LTMLE estimates over all combinations of all potential adjustment sets at each time-point. This means that employing LGBCEE can be computationally intensive when the number of variables or the number of times-point is large. 
Secondly, we focused on a setting where the outcome is measured at the end of the study. MSMs have also long been used for repeated outcome measures \citep{hernan2002estimating} and for time-to-event outcomes with overlapping treatment and outcome observation periods \citep{hernan2000marginal}. The methods proposed in this article could be extended to such cases. A simple way of achieving this with repeated outcome measures would be to repeatedly apply LGBCEE for each of the outcome measurement times. However, the computational burden of this approach might be excessive. Thirdly, LGBCEE is based on the assumption that exposure models are correctly specified to accurately identify the determinants of exposure. Incorrect specification of these models can lead to sub-optimal selection of covariates.
Extensions incorporating flexible machine learning approaches to exposure
modeling could mitigate this limitation. Furthermore, while LGBCEE model accounts for observed confounders, it does not directly address unmeasured confounders. A recent study has shown that machine learning algorithms can take advantage of the large number of variables available in medical administrative databases to find proxies for unmeasured confounders \citep{benasseur2022comparison}. Future work could apply LGBCEE on a database with a large quantity of variables to detect its ability to select proxies for unmeasured confounders. However, more scalable implementations adapted to the high-dimensional context would first need to be developed. 

In summary, LGBCEE represents an important methodological advance for estimating longitudinal causal effects. Its application to the study of the effects of statins on CVD risk in older adults has confirmed its practical utility. With the increased availability of longitudinal health data, notably via administrative databases and observational cohorts, LGBCEE represents a promising solution for improving the precision of causal estimation. 

\section*{Acknowledgements}
This work was supported by a Canadian Institutes of Health Research grant (\#420060). DT and JRG were supported by research career awards from the Fonds de recherche du Qu\'ebec -- Sant\'e (\url{https://doi.org/10.69777/312198} and \url{https://doi.org/10.69777/330678}, respectively). MES is supported by a Canada Research Chair tier 2 in Causal Inference and Machine Learning in Health Science. The authors would like to thank Genevi\`eve Lefebvre for comments on a previous version of this manuscript. Real-data illustration based on data from the \copyright Government of Quebec (2024). The Government of Quebec is not responsible for the results produced using their data nor for their interpretation.

\newpage
\bibliographystyle{unsrt}  
\bibliography{references}  

\newpage

\appendix

{\center \huge  Supplementary material for ``A Bayesian model averaging approach for estimating longitudinal treatment effects'' by David Adenyo, Mireille E Schnitzer, Jason R Guertin and Denis Talbot}

\vspace{3cm}

\section{Detailed workflow example}
\noindent In this section we present a workflow example that details the different steps involved in LGBCEE. 
For this purpose we generated a sample of size $n= 500$ satisfying the following relationships:
	\begin{eqnarray*} 
		\mbox{logit}\{P(A_0=1)\}
		&=& -0.15 + X_{01} \\
             \mbox{logit}\{P(A_1=1)\}
		&=& -0.25 + A_{0}+ 2X_{11}   \\
  	   Y &=& -A_{1}+0.1X_{11}+\epsilon_Y   
	\end{eqnarray*}
 with $X_{01} \sim N(0,1)$, $X_{11} \sim N(A_0,1)$ and $\epsilon_Y \sim N(0,1)$, all independent. 
 
 We present a detailed calculation of the posterior probability ${P}(\alpha^{Y} = (0, 0, 1),{\alpha^{Q_1} = 0}|\mbox{O})$, where the elements of $\alpha^{Y}$ indicate the inclusion of $X_{01}$, $A_0$ and $X_{11}$ in the model for $Y$, respectively and $\alpha^{Q_1}$ indicate the inclusion of $X_{01}$ in the model for $Q_{1}$. Because this adjustment set meets the variable selection goal described in Section 3 of the main text, it is desirable that it has a large posterior probability.  

The first step of LGBCEE is to calculate the posterior distribution of the treatment model at each time-point, 
${P}(\alpha^{A_0}|{X}_{01},{A}_{0})\propto {P}(A_0|\alpha^{A_0},{X}_{01})$ and ${P}(\alpha^{A_1}
 |{X}_{01},A_0,{X}_{11},{A}_{1})\propto {P}(A_1|\alpha^{A_1},{X}_{01},A_0,{X}_{11})$.  
The two possible exposure models for $A_0$ in this example are:
	\begin{eqnarray*} 
		{A}_{0}|\longrightarrow  \alpha^{A_0}=(\alpha_{01}^{{A}_{0}}=0),  \\	
		 {A}_{0}|{X}_{01}\longrightarrow   \alpha^{A_0}=(\alpha_{01}^{{A}_{0}}=1).
	\end{eqnarray*} 
	We approximate ${P}(A_0|\alpha^{A_0},{X}_{01})$ using $\mbox{exp}[-0.5BIC(\alpha^{A_0})]$ \citep{Clyde2003}, where $BIC(\alpha^{A_0})$ is the Bayesian information criterion for the exposure model $\alpha^{A_0}$. In our example, model $\alpha^{A_0}=1$ received all posterior weight in the simulated dataset, that is
${{P}({\alpha^{A_0}} = 1|{X}_{01},A_0)=1}$. 

 The eight possible $A_1$ exposure models in this example are:
	\begin{eqnarray*} 
		{A}_{1}|\longrightarrow  \alpha^{A_1}=(\alpha_{11}^{{A}_{1}}=0,\alpha_{12}^{{A}_{1}}=0,\alpha_{13}^{{A}_{1}}=0),  \\
  	{A}_{1}|{X}_{11}\longrightarrow  \alpha^{A_1}=(\alpha_{11}^{{A}_{1}}=0,\alpha_{12}^{{A}_{1}}=0,\alpha_{13}^{{A}_{1}}=1),  \\
		{A}_{1}|{X}_{01}\longrightarrow  \alpha^{A_1}=(\alpha_{11}^{{A}_{1}}=1,\alpha_{12}^{{A}_{1}}=0,\alpha_{13}^{{A}_{1}}=0),  \\
		{A}_{1}|{X}_{01},{X}_{11}\longrightarrow \alpha^{A_1}=(\alpha_{11}^{{A}_{1}}=1,\alpha_{12}^{{A}_{1}}=0,\alpha_{13}^{{A}_{1}}=1),  \\
	{A}_{1}|{A}_{0}\longrightarrow  \alpha^{A_1}=(\alpha_{11}^{{A}_{1}}=0,\alpha_{12}^{{A}_{1}}=1,\alpha_{13}^{{A}_{1}}=0),  \\
 {A}_{1}|{A}_{0},{X}_{11}\longrightarrow  \alpha^{A_1}=(\alpha_{11}^{{A}_{1}}=0,\alpha_{12}^{{A}_{1}}=1,\alpha_{13}^{{A}_{1}}=1),  \\
{A}_{1}|{X}_{01},{A}_{0}\longrightarrow  \alpha^{A_1}=(\alpha_{11}^{{A}_{1}}=1,\alpha_{12}^{{A}_{1}}=1,\alpha_{13}^{{A}_{1}}=0),  \\
{A}_{1}|{X}_{01},{A}_{0},{X}_{11}\longrightarrow  \alpha^{A_1}=(\alpha_{11}^{{A}_{1}}=1,\alpha_{12}^{{A}_{1}}=1,\alpha_{13}^{{A}_{1}}=1).
	\end{eqnarray*}
Again, we approximate $P(A_1|\alpha^{A_1},{X}_{01},A_0,{X}_{11})$ using $\mbox{exp}[-0.5BIC(\alpha^{A_1})]$ and find ${{P}(\alpha^{A_1}=(0,1,1)|A_0,{X}_{11}, A_1)=1}$ in the simulated dataset.

The second step is to compute the prior distribution of the outcome model using ${P}(\alpha^{Y}|{X}_{01},A_0,{X}_{11}, A_1)=\sum_{\alpha^{A_1}}{P}(\alpha^{Y}|\alpha^{A_1}){P}(\alpha^{A_1}
 |{X}_{01},A_0,{X}_{11}, A_1)={P}(\alpha^{Y}|{\alpha^{A_1}} = (0, 1, 1))$. We take $\omega=100\times \sqrt{n}$. The eight possible outcome models are:
	\begin{eqnarray*} 
	Y|{A}_{1}\longrightarrow  \alpha^{Y}=(\alpha_{11}^{Y}=0,\alpha_{12}^{Y}=0,\alpha_{13}^{Y}=0), \\
	Y|{X}_{11},{A}_{1}\longrightarrow  \alpha^{Y}=(\alpha_{11}^{Y}=0,\alpha_{12}^{Y}=0,\alpha_{13}^{Y}=1), \\
  Y|{X}_{01},{A}_{1}\longrightarrow  \alpha^{Y}=(\alpha_{11}^{Y}=1,\alpha_{12}^{Y}=0,\alpha_{13}^{Y}=0), \\
  Y|{X}_{01},{X}_{11},{A}_{1}\longrightarrow  \alpha^{Y}=(\alpha_{11}^{Y}=1,\alpha_{12}^{Y}=0,\alpha_{13}^{Y}=1), \\
 Y|{A}_{0},{A}_{1}\longrightarrow  \alpha^{Y}=(\alpha_{11}^{Y}=0,\alpha_{12}^{Y}=1,\alpha_{13}^{Y}=0), \\
  Y|{A}_{0},{X}_{11},{A}_{1}\longrightarrow  \alpha^{Y}=(\alpha_{11}^{{Y}}=0,\alpha_{12}^{Y}=1,\alpha_{13}^{Y}=1), \\
 Y|{X}_{01},{A}_{0},{A}_{1}\longrightarrow  \alpha^{Y}=(\alpha_{11}^{{Y}}=1,\alpha_{12}^{Y}=1,\alpha_{13}^{Y}=0), \\
 Y|{X}_{01},{A}_{0},{X}_{11},{A}_{1}\longrightarrow  \alpha^{Y}=(\alpha_{11}^{{Y}}=1,\alpha_{12}^{Y}=1,\alpha_{13}^{Y}=1). 
	\end{eqnarray*}
	\noindent In the following, we present the detailed calculation of ${P}(\alpha^{Y}|\alpha^{A_1} = (0, 1, 1))$ for model  $\alpha^{Y}=(0, 0, 1)$ only. We have
	${P}(\alpha^{Y}=(0,0,1))={P}(\alpha^{Y}=(0,0,1)|\alpha^{A_1}=(0,1,1))\propto {P}(\alpha_{11}^{Y}=0|\alpha_{12}^{Y}=0,\alpha_{13}^{Y}=1,\alpha_{11}^{A_1}=0){P}(\alpha_{12}^{Y}=0|\alpha_{11}^{Y}=0,\alpha_{13}^{Y}=1,\alpha_{12}^{A_1}=1){P}(\alpha_{13}^{Y}=1|\alpha_{11}^{Y}=0,\alpha_{12}^{Y}=0,\alpha_{13}^{A_1}=1)$, where
	\[
	{P}(\alpha_{11}^{Y}=0|\alpha_{12}^{Y}=0,\alpha_{13}^{Y}=1,\alpha_{11}^{A_1}=0)=\frac{1}{2}, \ \
	{P}(\alpha_{12}^{Y}=0|\alpha_{11}^{Y}=0,\alpha_{13}^{Y}=1,\alpha_{12}^{A_1}=1)=\frac{1}{\omega_{12}^{\alpha^{Y}}+1},\]
	\[{P}(\alpha_{13}^{Y}=1|\alpha_{11}^{Y}=0,\alpha_{12}^{Y}=0,\alpha_{13}^{A_1}=1)=\frac{\omega_{13}^{\alpha^{Y}}}{\omega_{13}^{\alpha^{Y}}+1}.
	\]
	Recall that $\omega_{12}^{\alpha^{Y}}=\omega(\hat{\tilde{\delta}}_{12}^{\alpha^{Y}}\times \sigma_{A_{0}}/\sigma_{Y})^{2}$. Because $A_{0}$ is not in $\alpha^{Y}=(0,0,1)$, we get the regression parameter estimate for $A_{0}$ from the outcome model that includes the same covariates as $\alpha^{Y}=(0,0,1)$, but additionally includes  $A_{0}$, that is $\hat{\tilde{\delta}}_{12}^{\alpha^{Y}=(0,0,1)}=\hat{\delta}_{12}^{\alpha^{Y}=(0,1,1)}$. Also, we have $\omega_{13}^{\alpha^{Y}}=\omega(\hat{\delta}_{13}^{\alpha^{Y}}\times \sigma_{X_{11}}/\sigma_{Y})^{2}$.  
Finally, the value of the (unnormalized) prior probability of model $\alpha^{Y}=(0,0,1)$ is $0.312$. Table \ref{tab:7.0a} reports some of the intermediary steps in calculating the prior probability of each $\alpha^Y$.

\begin{table}
  \centering
\caption{\centering Calculation of the LGBCEE outcome model posterior distribution at time $t = 1$ according to different $\alpha^{Y}$ adjustment sets.}
\scalebox{0.75}{
\begin{threeparttable}
\begin{tabular}{rrrrrrr}
\cmidrule{2-6}          & \multicolumn{1}{c}{Model} & \multicolumn{1}{c}{BIC($\alpha^{Y}$)} &  \multicolumn{1}{c}{$P(Y|\alpha^{Y},X_{01},A_0,X_{11},A_1)$} & \multicolumn{1}{c}{${P}(\alpha^{Y}|X_{01},A_0,X_{11},A_1)$} & \multicolumn{1}{c}{$Q_1$} &  \\
\cmidrule{2-6}  & \multicolumn{1}{c}{$\alpha^{Y}=(0,0,0)$} & \multicolumn{1}{c}{1454.76}   &\multicolumn{1}{c}{0.0762}   & \multicolumn{1}{c}{0.0047} & \multicolumn{1}{c}{$\mathbb{\hat{E}}(Y|A_1 = a_1)$} &  \\
          & \multicolumn{1}{c}{$\alpha^{Y}=(0,0,1)$}       & \multicolumn{1}{c}{1449.98}   &\multicolumn{1}{c}{0.8344}   & \multicolumn{1}{c}{0.2878} & \multicolumn{1}{c}{$\mathbb{\hat{E}}(Y|X_{11},A_1=a_1)$} &  \\
          & \multicolumn{1}{c}{$\alpha^{Y}=(1,0,0)$}       & \multicolumn{1}{c}{1460.95}   &\multicolumn{1}{c}{0.0034}   & \multicolumn{1}{c}{0.0036} & \multicolumn{1}{c}{$\mathbb{\hat{E}}(Y|X_{01},A_1=a_1)$} &  \\
          & \multicolumn{1}{c}{$\alpha^{Y}=(1,0,1)$}       & \multicolumn{1}{c}{1456.04}   &\multicolumn{1}{c}{0.0402}   & \multicolumn{1}{c}{0.3612} & \multicolumn{1}{c}{$\mathbb{\hat{E}}(Y|X_{01},X_{11},A_1=a_1)$} &  \\
          & \multicolumn{1}{c}{$\alpha^{Y}=(0,1,0)$}       & \multicolumn{1}{c}{1460.90}   &\multicolumn{1}{c}{0.0035}   & \multicolumn{1}{c}{0.0015} & \multicolumn{1}{c}{$\mathbb{\hat{E}}(Y|A_0=a_0,A_1=a_1)$} &  \\
          & \multicolumn{1}{c}{$\alpha^{Y}=(0,1,1)$}       & \multicolumn{1}{c}{1456.04}   &\multicolumn{1}{c}{0.0402}   & \multicolumn{1}{c}{0.2060} & \multicolumn{1}{c}{$\mathbb{E}(Y|A_0=a_0,X_{11},A_1=a_1)$} &  \\
          & \multicolumn{1}{c}{$\alpha^{Y}=(1,1,0)$}       & \multicolumn{1}{c}{1467.04}   &\multicolumn{1}{c}{0.0002}   & \multicolumn{1}{c}{0.0025} & \multicolumn{1}{c}{$\mathbb{\hat{E}}(Y|X_{01},A_0=a_0,A_1=a_1)$} &  \\
          & \multicolumn{1}{c}{$\alpha^{Y}=(1,1,0)$}       & \multicolumn{1}{c}{1462.19}   &\multicolumn{1}{c}{0.0019}   & \multicolumn{1}{c}{0.1326} & \multicolumn{1}{c}{$\mathbb{\hat{E}}(Y|X_{01},A_0=a_0,X_{11},A_1=a_1)$} &  \\
\cmidrule{2-6}   
    \end{tabular}
 \begin{tablenotes}
      \small
      \item \textit{Note :} Each row corresponds to a candidate model characterized by the set $\alpha^Y$, indicating the inclusion of variables $X_{01}$, $A_0$ and $X_{11}$ in the adjustment model for $Y$, $P(Y|\alpha^{Y},X_{01},A_0,X_{11},A_1)$ is the marginal likelihood, BIC is the Bayesian information criterion, ${P}(\alpha^{Y}|X_{01},A_0,X_{11},A_1)$ and $Q_1$ is the pseudo-outcome associated with the $\alpha^Y$ model. The best model is the one with the lowest BIC value.
    \end{tablenotes}
  \end{threeparttable}}
      \label{tab:7.0a}
\end{table}

Next, we estimate the pseudo-outcome $Q_{1}=\mathbb{E}(Y|\alpha^{Y}=(0,0,1),X_{01},A_0,X_{11},A_1)$ by regressing $Y$ on ${A}_{1}$ and ${X}_{11}$ and then calculate $\hat{\mathbb{E}}(Y|X_{11},{A}_{1}={a}_{1})$ for each individual and each ${a}_1 \in \{0,1\}$. We then compute the prior distribution of the pseudo-outcome model, ${P}(\alpha^{Q_1}|\alpha^{Y}=(0,0,1))=\sum_{\alpha^{A_0}}{P}(\alpha^{Q_1}|\alpha^{Y}=(0,0,1),\alpha^{A_0}){P}(\alpha^{A_0}|{X}_{01},A_0)={P}(\alpha^{Q_1}|\alpha^{Y}=(0,0,1),\alpha^{A_0}=1)$. Since we focus on the outcome model $\alpha^{Y}=(0,0,1)$, there are two possible pseudo-outcome models:
	\begin{eqnarray*} 
		\mathbb{\hat{E}}(Y|{X}_{11},{A}_{1}={a}_{1})|{A}_{0}\longrightarrow  (\alpha^{Y}=(0,0,1),\alpha^{Q_{1}}=(\alpha_{01}^{Q_{1}}=0)), \\
  	 \mathbb{\hat{E}}(Y|{X}_{11},{A}_{1}={a}_{1})|{X}_{01},{A}_{0}\longrightarrow  (\alpha^{Y}=(0,0,1),\alpha^{Q_{1}}=(\alpha_{01}^{Q_{1}}=1)).
	\end{eqnarray*} 
We present the calculation of ${P}(\alpha^{Q_1}|\alpha^{Y})$ for adjustment set  $(\alpha^{Y}=(0,0,1),\alpha^{Q_{1}}=0)$. We have:
${P}(\alpha^{Q_{1}}=0|\alpha^{Y}=(0,0,1))\propto {P}(\alpha_{01}^{{Q_{1}}}=0|\alpha^{Y}=(0,0,1),\alpha_{01}^{A_0}=1)=\dfrac{1}{\omega_{01}^{\alpha^{Q_{1}}}+1}$, with
$\omega_{01}^{\alpha^{Q_{1}}}=\omega(\hat{\tilde{\delta}}_{01}^{{\alpha^{Q_{1}}}}\times \sigma_{X_{01}}/\sigma_{Q_{1}})^{2}$ and $\hat{\tilde{\delta}}_{01}^{{\alpha^{Q_{1}}}}=\hat{{\delta}}_{01}^{{\alpha^{Q_{1}}}}$. 
The value of the (unnormalized) prior probability of model $\alpha^{Q_{1}}$ is $0.026$. Table \ref{tab:7.0b} presents some of the intermediary steps in calculating the prior probability of each $\alpha^{Q_1}$.

\begin{table}
  \centering
  \caption{\centering Calculation of the LGBCEE outcome model posterior distribution at time $t = 0$ according to different $\alpha^{Q_1}$ adjustment sets conditional on covariates included in $\alpha^{Y}$.}
\begin{threeparttable}
    \begin{tabular}{rrrrrr}
\cmidrule{2-5}          & \multicolumn{1}{c}{Model} & \multicolumn{1}{c}{$BIC(\alpha^{Q_1}|\alpha^{Y})$}   &\multicolumn{1}{c}{$P(Q_{1}|\alpha^{Y},\alpha^{Q_{1}},A_0,X_{01})$} & \multicolumn{1}{c}{$P(\alpha^{Q_{1}}|\alpha^{Y},A_0,X_{01})$} &  \\
\cmidrule{2-5}
          & \multicolumn{1}{c}{$\alpha^{Q_1}=0|\alpha^{Y}=(0,0,0)$} &\multicolumn{1}{c}{647.16}  &\multicolumn{1}{c}{0.9693} & \multicolumn{1}{c}{0.1250} &  \\
          & \multicolumn{1}{c}{$\alpha^{Q_1}=1|\alpha^{Y}=(0,0,0)$} &\multicolumn{1}{c}{654.07}  &\multicolumn{1}{c}{0.0307} & \multicolumn{1}{c}{0.0000} &  \\
          & \multicolumn{1}{c}{$\alpha^{Q_1}=0|\alpha^{Y}=(0,0,1)$} &\multicolumn{1}{c}{1489.23} &\multicolumn{1}{c}{0.9693}  & \multicolumn{1}{c}{0.1245} &  \\
          & \multicolumn{1}{c}{$\alpha^{Q_1}=1|\alpha^{Y}=(0,0,1)$} &\multicolumn{1}{c}{1496.14} &\multicolumn{1}{c}{0.0307}  & \multicolumn{1}{c}{0.0005} &  \\
          & \multicolumn{1}{c}{$\alpha^{Q_1}=0|\alpha^{Y}=(1,0,0)$} &\multicolumn{1}{c}{639.99}  &\multicolumn{1}{c}{0.9639} & \multicolumn{1}{c}{0.0647} &  \\
          & \multicolumn{1}{c}{$\alpha^{Q_1}=1|\alpha^{Y}=(1,0,0)$} &\multicolumn{1}{c}{646.56}  &\multicolumn{1}{c}{0.0361} & \multicolumn{1}{c}{0.0603} &  \\
          & \multicolumn{1}{c}{$\alpha^{Q_1}=0|\alpha^{Y}=(1,0,1)$} &\multicolumn{1}{c}{1484.55} &\multicolumn{1}{c}{0.9521}  & \multicolumn{1}{c}{0.0357} &  \\
          & \multicolumn{1}{c}{$\alpha^{Q_1}=1|\alpha^{Y}=(1,0,0)$} &\multicolumn{1}{c}{1490.53} &\multicolumn{1}{c}{0.0479}  & \multicolumn{1}{c}{0.0893} &  \\
          & \multicolumn{1}{c}{$\alpha^{Q_1}=0|\alpha^{Y}=(0,1,0)$} &\multicolumn{1}{c}{678.17}  &\multicolumn{1}{c}{0.9693} & \multicolumn{1}{c}{0.1250} &  \\
          & \multicolumn{1}{c}{$\alpha^{Q_1}=1|\alpha^{Y}=(0,1,0)$} &\multicolumn{1}{c}{685.08}  &\multicolumn{1}{c}{0.0307} & \multicolumn{1}{c}{0.0000} &  \\
          & \multicolumn{1}{c}{$\alpha^{Q_1}=0|\alpha^{Y}=(0,1,1)$} &\multicolumn{1}{c}{1479.37} &\multicolumn{1}{c}{0.9693}  & \multicolumn{1}{c}{0.1245} &  \\
          & \multicolumn{1}{c}{$\alpha^{Q_1}=1|\alpha^{Y}=(0,1,1)$} &\multicolumn{1}{c}{1486.27} &\multicolumn{1}{c}{0.0307}  & \multicolumn{1}{c}{0.0005} &  \\
          & \multicolumn{1}{c}{$\alpha^{Q_1}=0|\alpha^{Y}=(1,1,0)$} &\multicolumn{1}{c}{678.24}  &\multicolumn{1}{c}{0.9402} & \multicolumn{1}{c}{0.0258} &  \\
          & \multicolumn{1}{c}{$\alpha^{Q_1}=1|\alpha^{Y}=(1,1,0)$} &\multicolumn{1}{c}{683.75}  &\multicolumn{1}{c}{0.0598} & \multicolumn{1}{c}{0.0992} &  \\
          & \multicolumn{1}{c}{$\alpha^{Q_1}=0|\alpha^{Y}=(1,1,1)$} &\multicolumn{1}{c}{1478.63} &\multicolumn{1}{c}{0.9617}  & \multicolumn{1}{c}{0.0554} &  \\
          & \multicolumn{1}{c}{$\alpha^{Q_1}=1|\alpha^{Y}=(1,1,1)$} &\multicolumn{1}{c}{1485.07} &\multicolumn{1}{c}{0.0383}  & \multicolumn{1}{c}{0.0696} &  \\
\cmidrule{2-5}   
    \end{tabular}%
 \begin{tablenotes}
      \small
      \item \textit{Note :} Each row corresponds to a candidate model characterized by the set $\alpha^{Q_1}$, indicating the inclusion of variables $X_{01}$ in the adjustment model for $Q_1$ conditional on covariates included in $\alpha^{Y}$, 
      $P(Q_{1}|\alpha^{Y},\alpha^{Q_{1}},A_0,X_{01})$ is the marginal likelihood conditional on covariates included in $\alpha^{Y}$, 
      BIC is the Bayesian information criterion conditional on covariates included in $\alpha^{Y}$, 
      $P(\alpha^{Q_{1}}|\alpha^{Y},A_0,X_{01})$ is the posterior probability conditional on covariates included in $\alpha^{Y}$.
    \end{tablenotes}
  \end{threeparttable}
    \label{tab:7.0b}
\end{table}

Following the same process for the fifteen other adjustment sets ($\alpha^{Y}, \alpha^{Q_1}$), we calculate the unnormalized prior probabilities and then normalize them such that their sum equals 1 (see the second column of Table \ref{tab:7.1}).  We finally calculate the joint posterior distribution of the adjustment set ($\alpha^{Y}=(0,0,1),\alpha^{Q_{1}}=0$) using the relationship $P((\alpha^{Y}=(0,0,1),\alpha^{Q_{1}}=0)|\mbox{O})\propto P(Y|\alpha^{Y}=(0,0,1),X_{11},A_1)P(Q_{1}|\alpha^{Y}=(0,0,1),\alpha^{Q_{1}}=0,A_0)P(\alpha^{Y}=(0,0,1))P(\alpha^{Q_{1}}=0|\alpha^{Y}=(0,0,1))$. Again, we use $\mbox{exp}[-0.5BIC(\alpha^{Y}=(0,0,1))]$ and $\mbox{exp}[-0.5BIC(\alpha^{Q_{1}}=0)]$ to approximate $P(Y|\alpha^{Y}=(0,0,1),X_{11},A_1)$ and $P(Q_{1}|\alpha^{Y}=(0,0,1),\alpha^{Q_{1}}=0,A_0)$, respectively. Table \ref{tab:7.1} provides the results of the intermediate steps. 

We can see that if one were to use a noninformative prior ${P}(\alpha^{Y}|X_{01},A_0,X_{11},A_1){P}(\alpha^{Q_1}|\alpha^{Y},X_{01},A_0) \propto 1$, the bulk of the posterior weight would go to a pair of models that does not satisfy conditional exchangeability, whereas the proposed prior succeeded in redirecting the bulk of the posterior weight onto sufficient adjustment sets. In addition, the adjustment set with the most posterior weight is the one that conforms best with our variable selection objective.

\begin{table}
  \centering
  \caption{\centering Calculation of the LGBCEE outcome model posterior distribution with intermediate steps.}
      \begin{threeparttable}
    \begin{tabular}{rrrrrr} 
\cmidrule{2-5}          & \multicolumn{1}{c}{($\alpha^{Y},\alpha^{Q_1}$)} & \multicolumn{1}{c}{LGBCEE prior} & \multicolumn{1}{c}{$P(\alpha^{Y},\alpha^{Q_{1}}|\mbox{O})$} & \multicolumn{1}{c}{$P(\alpha^{Y},\alpha^{Q_{1}}|\mbox{O})$} &  \\
& & &  w/ n-inf prior & w/ LGBCEE prior \\
\cmidrule{2-5}          & \multicolumn{1}{l}{((0,0,0),0)} & \multicolumn{1}{l}{0.0122} & \multicolumn{1}{c}{0.6161} & \multicolumn{1}{c}{0.0767} &  \\
          & \multicolumn{1}{l}{((0,0,0),1)} & \multicolumn{1}{l}{0.0000} & \multicolumn{1}{c}{0.0195} & \multicolumn{1}{c}{0.0000} &  \\
          & \multicolumn{1}{l}{((0,0,1),0)} & \multicolumn{1}{l}{0.3583} & \multicolumn{1}{c}{0.2254} & \multicolumn{1}{c}{0.8259} &  \\
          & \multicolumn{1}{l}{((0,0,1),1)} & \multicolumn{1}{l}{0.0189} & \multicolumn{1}{c}{0.0072} & \multicolumn{1}{c}{0.0014} &  \\
          & \multicolumn{1}{l}{((1,0,0),0)} & \multicolumn{1}{l}{0.0001} & \multicolumn{1}{c}{0.0055} & \multicolumn{1}{c}{0.0000} &  \\
          & \multicolumn{1}{l}{((1,0,0),1)} & \multicolumn{1}{l}{0.0030} & \multicolumn{1}{c}{0.0500} & \multicolumn{1}{c}{0.0015} &  \\
          & \multicolumn{1}{l}{((1,0,1),0)} & \multicolumn{1}{l}{0.0092} & \multicolumn{1}{c}{0.0029} & \multicolumn{1}{c}{0.0003} &  \\
          & \multicolumn{1}{l}{((1,0,1),1)} & \multicolumn{1}{l}{0.2848} & \multicolumn{1}{c}{0.0270} & \multicolumn{1}{c}{0.0787} &  \\
          & \multicolumn{1}{l}{((0,1,0),0)} & \multicolumn{1}{l}{0.0066} & \multicolumn{1}{c}{0.0295} & \multicolumn{1}{c}{0.0020} &  \\
          & \multicolumn{1}{l}{((0,1,0),1)} & \multicolumn{1}{l}{0.0000} & \multicolumn{1}{c}{0.0009} & \multicolumn{1}{c}{0.0000} &  \\
          & \multicolumn{1}{l}{((0,1,1),0)} & \multicolumn{1}{l}{0.0980} & \multicolumn{1}{c}{0.0104} & \multicolumn{1}{c}{0.0104} &  \\
          & \multicolumn{1}{l}{((0,1,1),1)} & \multicolumn{1}{l}{0.0056} & \multicolumn{1}{c}{0.0003} & \multicolumn{1}{c}{0.0000} &  \\
          & \multicolumn{1}{l}{((1,1,0),0)} & \multicolumn{1}{l}{0.0003} & \multicolumn{1}{c}{0.0000} & \multicolumn{1}{c}{0.0000} &  \\
          & \multicolumn{1}{l}{((1,1,0),1)} & \multicolumn{1}{l}{0.0139} & \multicolumn{1}{c}{0.0037} & \multicolumn{1}{c}{0.0005} &  \\
          & \multicolumn{1}{l}{((1,1,1),0)} & \multicolumn{1}{l}{0.0049} & \multicolumn{1}{c}{0.0000} & \multicolumn{1}{c}{0.0000} &  \\
          & \multicolumn{1}{l}{((1,1,1),1)} & \multicolumn{1}{l}{0.1843} & \multicolumn{1}{c}{0.0014} & \multicolumn{1}{c}{0.0026} &  \\
\cmidrule{2-5}
    \end{tabular}
    \begin{tablenotes}
      \small
      \item \textit{Note : } LGBCEE prior is the product of the prior distribution at each time-point ${P}(\alpha^{Y}|X_{01},A_0,X_{11},A_1){P}(\alpha^{Q_1}|\alpha^{Y},X_{01},A_0)$, $P(\alpha^{Y},\alpha^{Q_{1}}|\mbox{O})$ w/n-inf prior and 
      $P(\alpha^{Y},\alpha^{Q_{1}}|\mbox{O})$ w/LGBCEE prior are posterior probabilities according to a non-informative prior and the LGBCEE prior, respectively.
    \end{tablenotes}
  \end{threeparttable}
  \label{tab:7.1}
\end{table}

\clearpage

\section{Additional information on simulation scenarios}
\subsection{Scenarios \label{A6.1.1}}
Scenario 1 (\textit{Simple setting}): 
$C_{0}, P_{0}, I_{0},U_{0}\sim  N(0,1)$, all independent, $A_{0} \sim Bern(p_{A_{0}}=expit(C_{0}+I_{0}))$, $I_{1}\sim  N(0.5I_{0},1)$, $C_{1}\sim  N(C_0+0.5I_{0}+0.5I_{1} + 0.5A_0,1)$, $P_{1}\sim  N(P_0,1)$,
$U_{1}\sim  N(0,1)$, $A_{1} \sim Bern(p_{A_{1}}=expit(A_0+I_{0}+I_{1}+C_{1}))$ 
and $Y \sim N(-2-0.5A_{0}-0.5A_{1}+C_{0}+C_1 + 0.5P_{0}+0.5P_{1},1)$

The equation used for generating $Y$  to the binary outcome case (Scenarios 1B) is $Y \sim Bern(p_{Y}=expit(-2-0.5A_{0}-0.5A_{1}+C_{0}+C_1 + 0.5P_{0}+0.5P_{1}))$.

Scenario 2 (\textit{Conditional instruments}): 
The data-generating equations are $X_{04},...,X_{020} \sim  N(0,1)$, $X_{01},X_{02},X_{03} \sim  N(X_{06}+X_{07}+X_{08},1)$, $A_{0} \sim Bern(p_{A_{0}}=expit(X_{06}+\cdots+X_{015}))$, $X_{14},...,X_{120} \sim  N(0,1)$, $X_{11},X_{12},X_{13} \sim  N(A_0+X_{16}+X_{17}+X_{18} +X_{01}+X_{02}+X_{03},1)$, $A_{1} \sim Bern(p_{A_{1}}=expit(A_0+X_{16}+\cdots+X_{115}))$ and $Y \sim N(A_{0}+A_{1}+0.1\sum_{i=1}^{5}X_{0i}+0.1\sum_{i=1}^{5}X_{1i},1)$

Scenario 3 (\textit{Correlated covariates}): 
$X_{01},...,X_{040}$ were generated as multivariate normal variables with mean 0, unit variance, and 0.2 correlation, 
\[
A_{0} \sim Bern(p_{A_{0}}=expit(0.1\sum_{i=1}^{5}X_{0i}-0.1\sum_{i=6}^{10}X_{0i}+
0.1\sum_{i=11}^{14}X_{0i}-0.1\sum_{i=15}^{20}X_{0i})),
\]
\[
X_{11},\cdots,X_{15} \sim N(0.2\sum_{i=1}^{10}X_{0i}+0.05\sum_{i=11}^{20}X_{0i}+0.5A_0, 0.5^2),
\]
\[
X_{16}, X_{17} \sim N(0.2\sum_{i=11}^{20}X_{0i}+0.5A_0, 0.5^2), X_{18}, X_{19}, X_{110} \sim N(0.05\sum_{i=11}^{20}X_{0i}+0.2\sum_{i=21}^{30}X_{0i}, 0.5^2),
\]
\[
A_{1} \sim Bern(p_{A_{1}}=expit(0.1\sum_{i=1}^{5}X_{0i}-0.1\sum_{i=6}^{10}X_{0i}+
0.1\sum_{i=11}^{14}X_{0i}-0.1\sum_{i=15}^{20}X_{0i}+0.1\sum_{i=41}^{47}X_{1i}))
\]
and 
$Y \sim N(0.5A_{0}+0.5A_{1}+0.1\sum_{i=1}^{10}X_{0i}+0.02\sum_{i=11}^{20}X_{0i}
+0.1\sum_{i=21}^{30}X_{0i}+0.1\sum_{i=41}^{45}X_{1i}
+0.02X_{146}+0.02X_{147}+0.1\sum_{i=48}^{50}X_{1i}+0.1A_0X_{01}-0.05A_1X_{12},1)$
 

Scenario 4 (Nonlinear outcome model): 
At baseline ($t=0$), $X_{01}, \dots, X_{05}$ were generated as multivariate normal variables with mean $1$, unit variance, and pairwise correlation $0.6$. We then computed a quadratic term 
\[
U_0 = \mathrm{vec}(X_0 X_0^\top).
\]
The initial treatment 
\[
A_0 \sim \text{Bernoulli}\big(p = \text{expit}(0.5 X_{01} + 0.5 X_{02} + 0.1 X_{03})\big).
\]
At time 1 ($t=1$), covariates $X_1$ were generated as
\[
X_1 = -0.8 A_0 + 0.5 X_0 + \varepsilon,\quad \varepsilon \sim \text{MVN}(\mathbf{1}, \Sigma),
\]
where $\Sigma$ has unit variances and $0.6$ correlations. We computed 
\[
U_1 = \mathrm{vec}(X_1 X_1^\top).
\]
The second treatment 
\[
A_1 \sim \text{Bernoulli}\big(p = \text{expit}(0.5 X_{11} + 0.5 X_{12} + 0.1 X_{13})\big).
\]
Finally, the outcome 
\[
Y \sim N\Big(\big(-4 - 0.4 A_0 - 0.5 A_1 + 0.5 \sum_{i=3}^5 X_{0i} + 0.25 \sum U_0 + \sum_{i=3}^5 X_{1i} + 0.5 \sum U_1\big),1\Big)
\]
The equation used for generating $Y$  to the binary outcome case (Scenarios 4B) is 
\[
Y \sim \text{Bernoulli}\Big(p = \text{expit}\big(-4 - 0.4 A_0 - 0.5 A_1 + 0.5 \sum_{i=3}^5 X_{0i} + 0.25 \sum U_0 + \sum_{i=3}^5 X_{1i} + 0.5 \sum U_1\big)\Big).
\]

\clearpage

\subsection{Additional results \label{A6.1.2}}

\begin{table}[h]
  \centering
   \caption{Scenario 1 simulation study results for binary outcome}
  \scalebox{0.8}{
    \begin{tabular}{rrrrrr}
    \midrule
    \multicolumn{6}{c}{\textbf{$n = 250$}} \\
    \midrule
    \multicolumn{1}{l}{\textbf{Method}} & \multicolumn{1}{c}{\textbf{Bias}} & \multicolumn{1}{c}{\textbf{SD}} & \multicolumn{1}{c}{\textbf{ESE}} & \multicolumn{1}{c}{\textbf{RMSE}} & \multicolumn{1}{c}{\textbf{CP}} \\
    \multicolumn{1}{l}{LGBCEE} & \multicolumn{1}{c}{0.002} & \multicolumn{1}{c}{0.040} & \multicolumn{1}{c}{0.039} & \multicolumn{1}{c}{0.040} & \multicolumn{1}{c}{0.937} \\
    \multicolumn{1}{l}{Target-DR} & \multicolumn{1}{c}{-0.001} & \multicolumn{1}{c}{0.036} & \multicolumn{1}{c}{0.034} & \multicolumn{1}{c}{0.036} & \multicolumn{1}{c}{0.922} \\
    \multicolumn{1}{l}{Full-DR} & \multicolumn{1}{c}{0.000} & \multicolumn{1}{c}{0.046} & \multicolumn{1}{c}{0.039} & \multicolumn{1}{c}{0.046} & \multicolumn{1}{c}{0.909} \\
    \multicolumn{1}{l}{Target G-COMP} & \multicolumn{1}{c}{-0.001} & \multicolumn{1}{c}{0.032} & \multicolumn{1}{c}{0.032} & \multicolumn{1}{c}{0.032} & \multicolumn{1}{c}{0.935} \\
    \multicolumn{1}{l}{Full G-COMP} & \multicolumn{1}{c}{0.000} & \multicolumn{1}{c}{0.034} & \multicolumn{1}{c}{0.034} & \multicolumn{1}{c}{0.034} & \multicolumn{1}{c}{0.938} \\
    \multicolumn{1}{l}{IPTW} & \multicolumn{1}{c}{0.013} & \multicolumn{1}{c}{0.062} & \multicolumn{1}{c}{0.052} & \multicolumn{1}{c}{0.064} & \multicolumn{1}{c}{0.907} \\
    \multicolumn{1}{l}{C-LTMLE} & \multicolumn{1}{c}{0.021} & \multicolumn{1}{c}{0.044} & \multicolumn{1}{c}{0.043} & \multicolumn{1}{c}{0.049} & \multicolumn{1}{c}{0.891} \\
    \midrule
    \multicolumn{6}{c}{\textbf{$n = 500$}} \\
    \midrule
    \multicolumn{1}{l}{\textbf{Method}} & \multicolumn{1}{c}{\textbf{Bias}} & \multicolumn{1}{c}{\textbf{SD}} & \multicolumn{1}{c}{\textbf{ESE}} & \multicolumn{1}{c}{\textbf{RMSE}} & \multicolumn{1}{c}{\textbf{CP}} \\
    \multicolumn{1}{l}{LGBCEE} & \multicolumn{1}{c}{0.003} & \multicolumn{1}{c}{0.027} & \multicolumn{1}{c}{0.027} & \multicolumn{1}{c}{0.027} & \multicolumn{1}{c}{0.945} \\
    \multicolumn{1}{l}{Target-DR} & \multicolumn{1}{c}{0.001} & \multicolumn{1}{c}{0.025} & \multicolumn{1}{c}{0.024} & \multicolumn{1}{c}{0.025} & \multicolumn{1}{c}{0.935} \\
    \multicolumn{1}{l}{Full-DR} & \multicolumn{1}{c}{0.000} & \multicolumn{1}{c}{0.031} & \multicolumn{1}{c}{0.028} & \multicolumn{1}{c}{0.031} & \multicolumn{1}{c}{0.916} \\
    \multicolumn{1}{l}{Target G-COMP} & \multicolumn{1}{c}{0.000} & \multicolumn{1}{c}{0.023} & \multicolumn{1}{c}{0.023} & \multicolumn{1}{c}{0.023} & \multicolumn{1}{c}{0.937} \\
    \multicolumn{1}{l}{Full G-COMP} & \multicolumn{1}{c}{0.000} & \multicolumn{1}{c}{0.024} & \multicolumn{1}{c}{0.024} & \multicolumn{1}{c}{0.024} & \multicolumn{1}{c}{0.936} \\
    \multicolumn{1}{l}{IPTW} & \multicolumn{1}{c}{0.008} & \multicolumn{1}{c}{0.044} & \multicolumn{1}{c}{0.040} & \multicolumn{1}{c}{0.044} & \multicolumn{1}{c}{0.929} \\
    \multicolumn{1}{l}{C-LTMLE} & \multicolumn{1}{c}{0.025} & \multicolumn{1}{c}{0.031} & \multicolumn{1}{c}{0.030} & \multicolumn{1}{c}{0.040} & \multicolumn{1}{c}{0.835} \\
    \midrule
    \multicolumn{6}{c}{\textbf{$n = 1000$}} \\
    \midrule
    \multicolumn{1}{l}{\textbf{Method}} & \multicolumn{1}{c}{\textbf{Bias}} & \multicolumn{1}{c}{\textbf{SD}} & \multicolumn{1}{c}{\textbf{ESE}} & \multicolumn{1}{c}{\textbf{RMSE}} & \multicolumn{1}{c}{\textbf{CP}} \\
    \multicolumn{1}{l}{LGBCEE} & \multicolumn{1}{c}{0.002} & \multicolumn{1}{c}{0.019} & \multicolumn{1}{c}{0.019} & \multicolumn{1}{c}{0.019} & \multicolumn{1}{c}{0.947} \\
    \multicolumn{1}{l}{Target-DR} & \multicolumn{1}{c}{0.000} & \multicolumn{1}{c}{0.018} & \multicolumn{1}{c}{0.017} & \multicolumn{1}{c}{0.018} & \multicolumn{1}{c}{0.929} \\
    \multicolumn{1}{l}{Full-DR} & \multicolumn{1}{c}{-0.001} & \multicolumn{1}{c}{0.023} & \multicolumn{1}{c}{0.021} & \multicolumn{1}{c}{0.023} & \multicolumn{1}{c}{0.921} \\
    \multicolumn{1}{l}{Target G-COMP} & \multicolumn{1}{c}{0.000} & \multicolumn{1}{c}{0.016} & \multicolumn{1}{c}{0.016} & \multicolumn{1}{c}{0.016} & \multicolumn{1}{c}{0.936} \\
    \multicolumn{1}{l}{Full G-COMP} & \multicolumn{1}{c}{0.000} & \multicolumn{1}{c}{0.017} & \multicolumn{1}{c}{0.017} & \multicolumn{1}{c}{0.017} & \multicolumn{1}{c}{0.938} \\
    \multicolumn{1}{l}{IPTW} & \multicolumn{1}{c}{0.005} & \multicolumn{1}{c}{0.035} & \multicolumn{1}{c}{0.030} & \multicolumn{1}{c}{0.035} & \multicolumn{1}{c}{0.919} \\
    \multicolumn{1}{l}{C-LTMLE} & \multicolumn{1}{c}{0.027} & \multicolumn{1}{c}{0.023} & \multicolumn{1}{c}{0.022} & \multicolumn{1}{c}{0.035} & \multicolumn{1}{c}{0.724} \\
    \midrule
          &       &       &       &       &  \\
    \end{tabular}}
  \label{tab:addlabel}
\end{table}

\begin{table}[h]
   \centering
  \caption{Scenario 4 simulation study results for binary outcome}
  \scalebox{0.8}{
    \begin{tabular}{rrrrrrrr}
          & \multicolumn{6}{c}{}                          &  \\
\cmidrule{2-7}          & \multicolumn{6}{c}{$n = 250$}          &  \\
\cmidrule{2-7}          & \multicolumn{1}{l}{\textbf{Method}} & \multicolumn{1}{c}{\textbf{Bias}} & \multicolumn{1}{c}{\textbf{SD}} & \multicolumn{1}{c}{\textbf{ESE}} & \multicolumn{1}{c}{\textbf{RMSE}} & \multicolumn{1}{c}{\textbf{CP}} &  \\
          & \multicolumn{1}{l}{LGBCEE} & \multicolumn{1}{c}{-0.01} & \multicolumn{1}{c}{0.05} & \multicolumn{1}{c}{0.04} & \multicolumn{1}{c}{0.05} & \multicolumn{1}{c}{0.93} &  \\
          & \multicolumn{1}{l}{Target-DR} & \multicolumn{1}{c}{0.00} & \multicolumn{1}{c}{0.05} & \multicolumn{1}{c}{0.04} & \multicolumn{1}{c}{0.05} & \multicolumn{1}{c}{0.91} &  \\
          & \multicolumn{1}{l}{Full-DR} & \multicolumn{1}{c}{0.00} & \multicolumn{1}{c}{0.05} & \multicolumn{1}{c}{0.04} & \multicolumn{1}{c}{0.05} & \multicolumn{1}{c}{0.91} &  \\
          & \multicolumn{1}{l}{Target G-COMP} & \multicolumn{1}{c}{0.00} & \multicolumn{1}{c}{0.04} & \multicolumn{1}{c}{0.04} & \multicolumn{1}{c}{0.04} & \multicolumn{1}{c}{0.95} &  \\
          & \multicolumn{1}{l}{Full G-COMP} & \multicolumn{1}{c}{0.00} & \multicolumn{1}{c}{0.04} & \multicolumn{1}{c}{0.04} & \multicolumn{1}{c}{0.04} & \multicolumn{1}{c}{0.95} &  \\
          & \multicolumn{1}{l}{C-LTMLE} & \multicolumn{1}{c}{0.03} & \multicolumn{1}{c}{0.03} & \multicolumn{1}{c}{0.03} & \multicolumn{1}{c}{0.05} & \multicolumn{1}{c}{0.75} &  \\
\cmidrule{2-7}          & \multicolumn{6}{c}{$n = 500$}          &  \\
\cmidrule{2-7}          & \multicolumn{1}{l}{\textbf{Method}} & \multicolumn{1}{c}{\textbf{Bias}} & \multicolumn{1}{c}{\textbf{SD}} & \multicolumn{1}{c}{\textbf{ESE}} & \multicolumn{1}{c}{\textbf{RMSE}} & \multicolumn{1}{c}{\textbf{CP}} &  \\
          & \multicolumn{1}{l}{LGBCEE} & \multicolumn{1}{c}{0.00} & \multicolumn{1}{c}{0.03} & \multicolumn{1}{c}{0.03} & \multicolumn{1}{c}{0.03} & \multicolumn{1}{c}{0.95} &  \\
          & \multicolumn{1}{l}{Target-DR} & \multicolumn{1}{c}{0.00} & \multicolumn{1}{c}{0.04} & \multicolumn{1}{c}{0.03} & \multicolumn{1}{c}{0.04} & \multicolumn{1}{c}{0.94} &  \\
          & \multicolumn{1}{l}{Full-DR} & \multicolumn{1}{c}{0.00} & \multicolumn{1}{c}{0.04} & \multicolumn{1}{c}{0.03} & \multicolumn{1}{c}{0.04} & \multicolumn{1}{c}{0.94} &  \\
          & \multicolumn{1}{l}{Target G-COMP} & \multicolumn{1}{c}{0.00} & \multicolumn{1}{c}{0.03} & \multicolumn{1}{c}{0.03} & \multicolumn{1}{c}{0.03} & \multicolumn{1}{c}{0.94} &  \\
          & \multicolumn{1}{l}{Full G-COMP} & \multicolumn{1}{c}{0.00} & \multicolumn{1}{c}{0.03} & \multicolumn{1}{c}{0.03} & \multicolumn{1}{c}{0.03} & \multicolumn{1}{c}{0.94} &  \\
          & \multicolumn{1}{l}{C-LTMLE} & \multicolumn{1}{c}{0.03} & \multicolumn{1}{c}{0.02} & \multicolumn{1}{c}{0.02} & \multicolumn{1}{c}{0.04} & \multicolumn{1}{c}{0.71} &  \\
\cmidrule{2-7}          & \multicolumn{6}{c}{$n = 1000$}         &  \\
\cmidrule{2-7}          & \multicolumn{1}{l}{\textbf{Method}} & \multicolumn{1}{c}{\textbf{Bias}} & \multicolumn{1}{c}{\textbf{SD}} & \multicolumn{1}{c}{\textbf{ESE}} & \multicolumn{1}{c}{\textbf{RMSE}} & \multicolumn{1}{c}{\textbf{CP}} &  \\
          & \multicolumn{1}{l}{LGBCEE} & \multicolumn{1}{c}{0.00} & \multicolumn{1}{c}{0.02} & \multicolumn{1}{c}{0.03} & \multicolumn{1}{c}{0.02} & \multicolumn{1}{c}{0.95} &  \\
          & \multicolumn{1}{l}{Target-DR} & \multicolumn{1}{c}{0.00} & \multicolumn{1}{c}{0.03} & \multicolumn{1}{c}{0.03} & \multicolumn{1}{c}{0.03} & \multicolumn{1}{c}{0.94} &  \\
          & \multicolumn{1}{l}{Full-DR} & \multicolumn{1}{c}{0.00} & \multicolumn{1}{c}{0.03} & \multicolumn{1}{c}{0.03} & \multicolumn{1}{c}{0.03} & \multicolumn{1}{c}{0.94} &  \\
          & \multicolumn{1}{l}{Target G-COMP} & \multicolumn{1}{c}{0.00} & \multicolumn{1}{c}{0.02} & \multicolumn{1}{c}{0.02} & \multicolumn{1}{c}{0.02} & \multicolumn{1}{c}{0.93} &  \\
          & \multicolumn{1}{l}{Full G-COMP} & \multicolumn{1}{c}{0.00} & \multicolumn{1}{c}{0.02} & \multicolumn{1}{c}{0.02} & \multicolumn{1}{c}{0.02} & \multicolumn{1}{c}{0.93} &  \\
          & \multicolumn{1}{l}{C-LTMLE} & \multicolumn{1}{c}{0.03} & \multicolumn{1}{c}{0.02} & \multicolumn{1}{c}{0.02} & \multicolumn{1}{c}{0.03} & \multicolumn{1}{c}{0.57} &  \\
\cmidrule{2-7}   
    \end{tabular}}
  \label{tabA12}
\end{table}

\clearpage

\begin{figure}[p]
	\centerline{\includegraphics[width=12in,width=\textwidth]{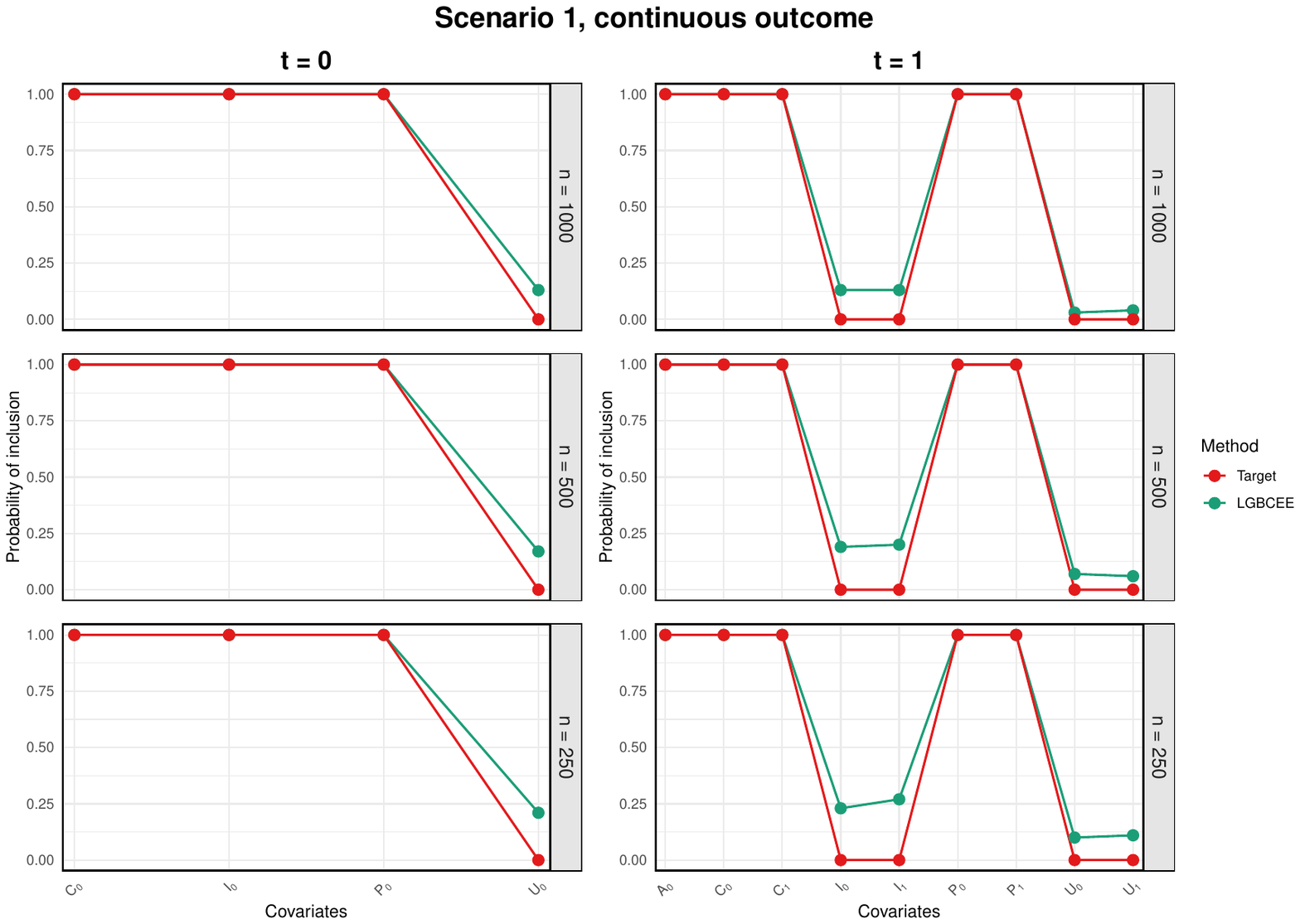}}
	\caption{Marginal probability of inclusion of covariates at each time-point.\label{fig4}}
\end{figure}
\begin{figure}[p]
	\centerline{\includegraphics[width=12in,width=\textwidth]{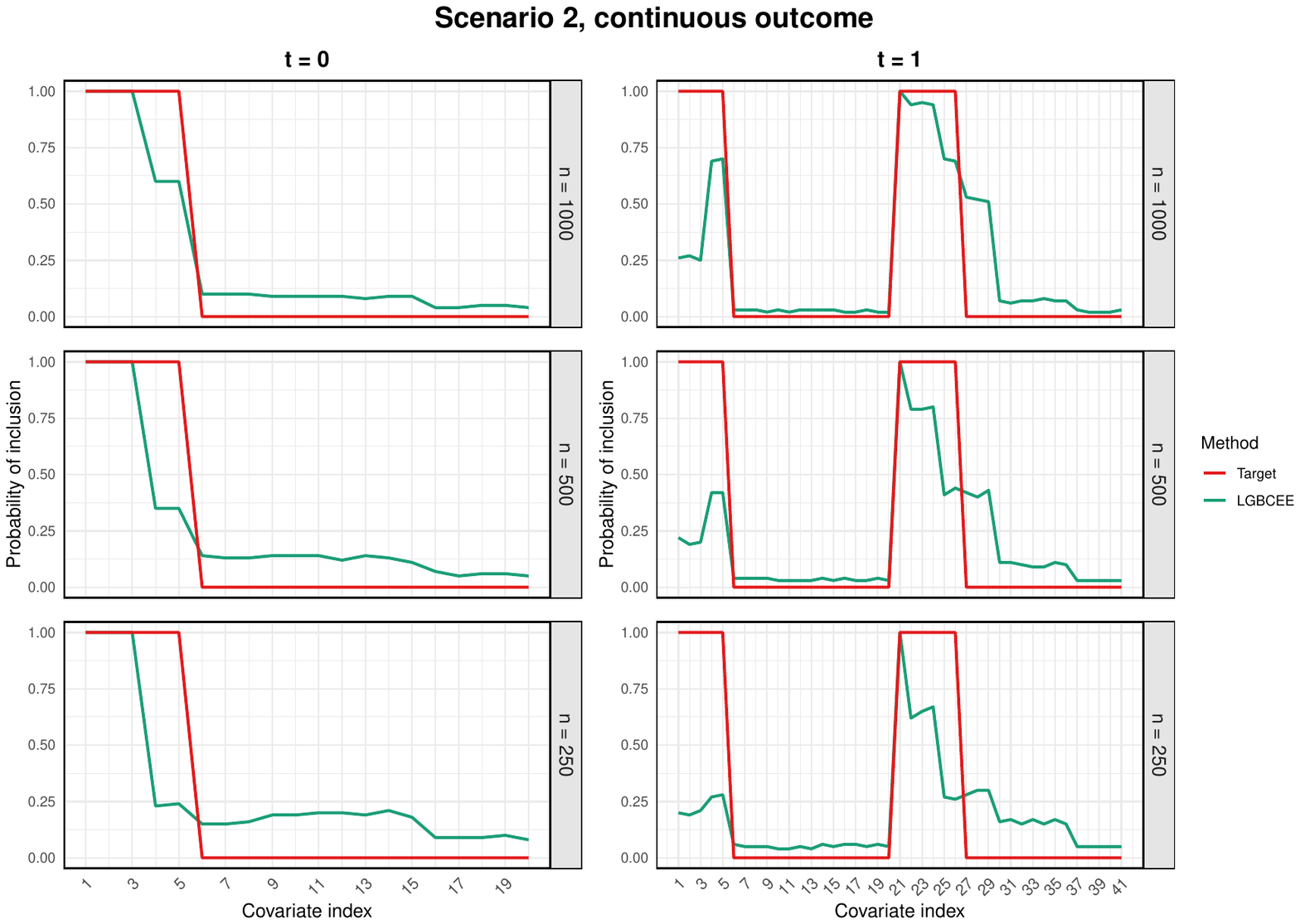}}
	\caption{Marginal probability of inclusion of covariates at each time-point.\label{fig5}}
\end{figure}
\begin{figure}[p]
	\centerline{\includegraphics[width=12in,width=\textwidth]{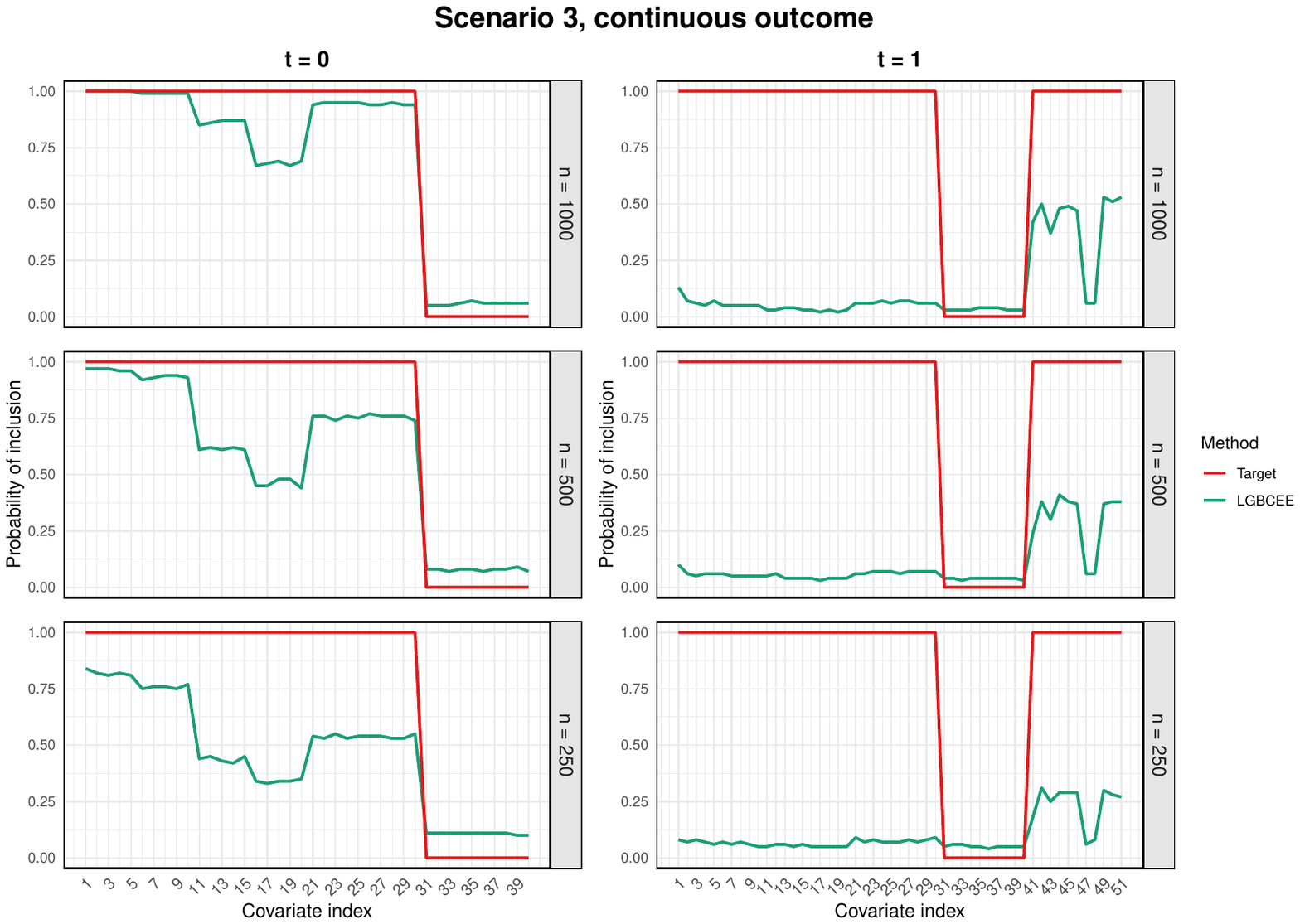}}
	\caption{Marginal probability of inclusion of covariates at each time-point.\label{fig6}}
\end{figure}
\begin{figure}[p]
	\centerline{\includegraphics[width=12in,width=\textwidth]{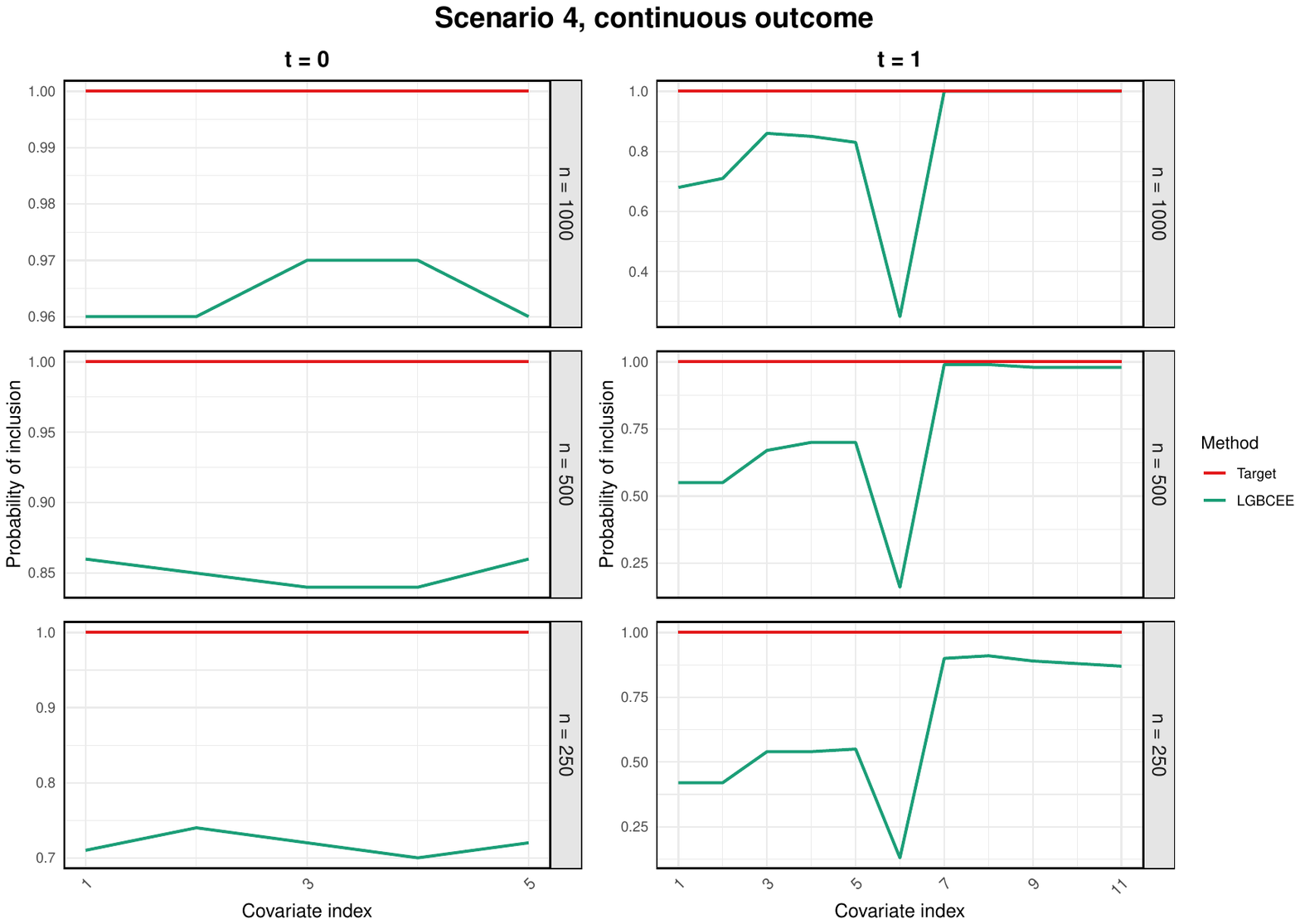}}
	\caption{Marginal probability of inclusion of covariates at each time-point.\label{fig7}}
\end{figure}
\begin{figure}[p]
	\centerline{\includegraphics[width=12in,width=\textwidth]{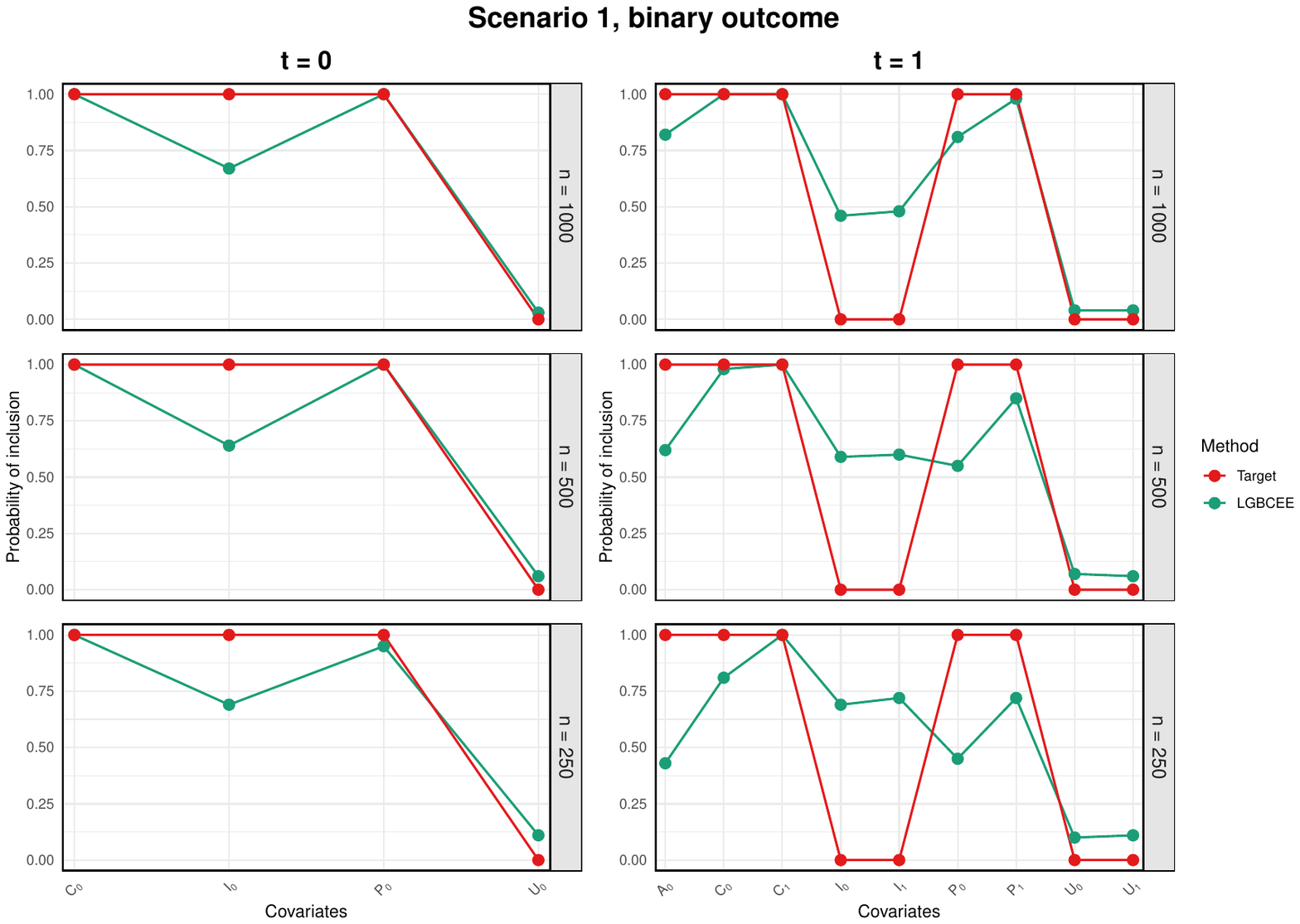}}
	\caption{Marginal probability of inclusion of covariates at each time-point.\label{fig8}}
\end{figure}
\begin{figure}[p]
	\centerline{\includegraphics[width=12in,width=\textwidth]{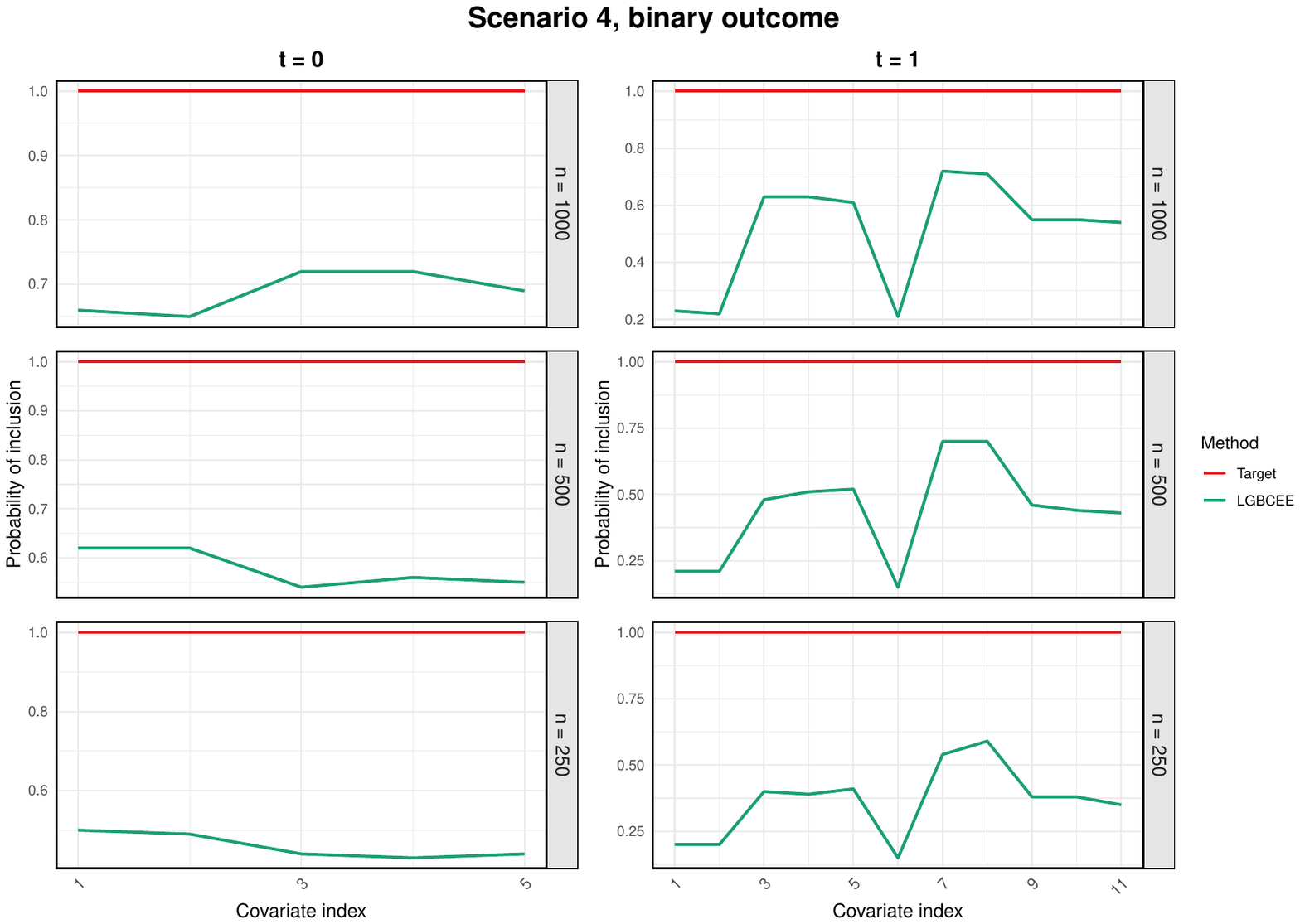}}
	\caption{Marginal probability of inclusion of covariates at each time-point.\label{fig9}}
\end{figure}

\clearpage

\section{The General Longitudinal Case of LGBCEE algorithm}
We now extend the above formulation to the general case with more than two time-points, i.e., $t=0,1,\dots,T$ with an arbitrary number of time-points $T\geq2$. In particular, a sample of independent and identically distributed observations $O=\{X_{i0},A_{i0},X_{i1},A_{i1},\dots,X_{iT},A_{iT},Y_i\}$, $i=1,\dots,n$, is drawn from our population where $X_{it}=\{X_{i,tn_t}; n_t=1,\dots,N_t\}$. 
To facilitate understanding, we denote 

\begin{align*}
{U}_{0}&=\{U_{01},\dots,U_{0N_0}\}, \ U_{0n_0} = X_{0n_0}, \ n_0 = 1,\dots,N_0, \\
{U}_{t}&= \{ U_{t1},\dots,U_{tN_0},U_{t(N_0+1)},U_{t(N_0+2)},\dots,U_{t(N_0+N_1+1)}, \\
&\qquad U_{t(N_0+N_1+2)},\dots,U_{t(N_0+N_1+\cdots+N_{t-1}+t)},  \\
&\qquad U_{t(N_0+N_1+\cdots+N_t+t)} \}, \quad t=1,\dots,T,
\end{align*}
where $U_{tn_0} = X_{0n_0}, n_0 = 1,\dots,N_0$, $U_{t(N_0+1)} = A_{0}$, 
$U_{t(N_0+n_1+1)} = X_{1n_1}, n_1 = 1,\dots,N_1$, $U_{t(N_0+N_1+2)} = A_{1}$,\dots,$U_{t(N_0+N_1+\cdots+N_{t-1}+t)} = A_{t-1}$, $U_{t(N_0+n_t+t)} = X_{tn_t}, n_t = 1,\dots,N_t$. In other words, $U_{t} \quad t=1,\dots,T$ represent all the variables available for adjustment at time-point $t$, respectively, including previous treatments. 

Let $\alpha^{Q_{t+1}}=(\alpha_{t1}^{Q_{t+1}},\dots,\alpha_{tN_0}^{Q_{t+1}},\alpha_{t(N_0+1)}^{Q_{t+1}},\alpha_{t(N_0+2)}^{Q_{t+1}},\dots,\alpha_{t(N_0+N_1+\cdots+N_t+t)}^{Q_{t+1}}), \ t=T,\dots,0$ be a vector of dimension $M_t=t+\sum_{k=0}^{t} N_k$ for the inclusion of the covariate history ${U}_{t}=\{X_0,A_0,X_1,A_1,\dots,A_{t-1},X_t\}$ in the potential adjustment set for $Q_{t+1}=\mathbb{E}[Q_{t+2}|\alpha^{Q_{t+2}},{A}_{t}={a}_{t},U_{t}]
$ with $Q_{T+1}\equiv Y$, where a component of $\alpha^{Q_{t+1}}$ is equal to $1$ if the variable associated with it is included in the set and this component equal to $0$ otherwise. Let $\underline{\alpha}^{Q}=(\alpha^{Q_{T+1}},\alpha^{Q_T},\dots,\alpha^{Q_1})$ and $\mathcal{A}_{t+1}, \ t=T,\dots,0$ be the set of all possible sets of confounders of $\alpha^{Q_{t+1}}$.  
Using this notation, LGBCEE estimates $\Delta$ as follows 
\[\hat{\Delta}\equiv \mathbb{E}[\Delta|\mbox{O}]=\sum_{\underline{\alpha}^{Q}\in \mathcal{A}_{T+1}\times \dots \times \mathcal{A}_{1}}\hat{\Delta}^{\underline{\alpha}^{Q}}{P}(\underline{\alpha}^{Q}|\mbox{O}),\] 
where $\hat{\Delta}^{\underline{\alpha}^{Q}}\equiv \mathbb{E}[\Delta|\mbox{O},\underline{\alpha}^{Q}]$ is the posterior mean of $\Delta$ after adjustment for the $\underline{\alpha}^{Q}$ variables. The weight assigned to each estimate $\hat{\Delta}^{\underline{\alpha}^{Q}}$ is the posterior probability, ${P}(\underline{\alpha}^{Q}|\mbox{O})\propto \prod_{t=0}^{T}P(Q_{t+1}|\alpha^{Q_{t+1}},A_t,U_t) {P}(\alpha^{Q_{t+1}}|\underline{\alpha}^{Q_{t+1}},A_t,U_t)$, where $\underline{\alpha}^{Q_{T+1}}=\emptyset$ and $\underline{\alpha}^{Q_{t+1}}=(\alpha^{Q_{t+2}},\dots,\alpha^{Q_{T+1}}), t=T-1,\dots 0$. Initializing $Q_{T+1}=Y$, we recursively estimate $Q_{t+1}$ for $t=T,\dots,0$ as
\[\begin{split} 
g\{\mathbb{E}(Q_{t+1}|\alpha^{Q_{t+1}},\underline{\alpha}^{Q_{t+1}},{A}_{t},{U}_{t})\}
=\delta_{t}^{\alpha^{Q_{t+1}}}+\beta_{t1}^{\alpha^{Q_{t+1}}}A_{t}+\sum_{m_t=1}^{M_t}\alpha_{tm_t}^{Q_{t+1}} \delta_{tm_t}^{\alpha^{Q_{t+1}}}U_{tm_t}.
\end{split} 
\]
We define $\alpha^{A_{t}}=(\alpha_{t1}^{A_{t}},\dots,\alpha_{tN_0}^{A_{t}},\alpha_{t(N_0+1)}^{A_{t}},\alpha_{t(N_0+2)}^{A_{t}},\dots,\alpha_{t(N_0+N_1+\dots+N_t+t)}^{A_{t}})\in \mathcal{A}_{t}$ similarly to $\alpha^{Q_{t+1}}$ i.e. $\alpha^{A_1}$ is a vector of dimension $M_t=t+\sum_{k=0}^{t} N_k$ for the inclusion of the history of covariates ${U}_{t}=\{X_0,A_0,X_1,A_1,\dots,A_{t-1},X_t\}$ in the exposure model at time $t$. The exposure models are given for $t=0,\dots,T$ by
\[\begin{split} 
\mbox{logit}[P(A_t=1|\alpha^{A_t},{U}_{t})]=\delta_{t}^{\alpha^{A_{t}}}+\sum_{m_t=1}^{M_t}\alpha_{tm_t}^{A_{t}} \delta_{tm_t}^{\alpha^{A_{t}}}U_{tm_t}.
\end{split} 
\]
The posterior distribution of the exposure model, ${P}(\alpha^{A_t}|{A}_{t},{U}_{t})\propto P(A_t|\alpha^{A_t},{A}_{t},{U}_{t}){P}(\alpha^{A_t})$, is then calculated. 

Similar to Section 3 of the main text, we make the following assumptions: (\textbf{B1}): $\alpha^{Q_{t+1}}$ is independent of (${A}_{t},{U}_{t}$) given ($\underline{\alpha}^{Q_{t+1}},\alpha^{A_t}$) and (\textbf{B2}): $\alpha^{A_t}$ is independent of $\underline{\alpha}^{Q_{t+1}}$ given (${A}_{t},{U}_{t}$). Finally the prior distribution of the LGBCEE is defined at each time-point as follows:
\begin{eqnarray*}
    {P}(\alpha^{Q_{t+1}}|\underline{\alpha}^{Q_{t+1}},{A}_{t},{U}_{t})
    &=& \sum_{\alpha^{A_t}\in \mathcal{A}_{t}} {P}(\alpha^{Q_{t+1}}|\underline{\alpha}^{Q_{t+1}},\alpha^{A_t},{A}_{t},{U}_{t}){P}(\alpha^{A_t}|\underline{\alpha}^{Q_{t+1}},{A}_{t},{U}_{t}) \\
    &\overset{\text{using (\textbf{B1}),(\textbf{B2})}}{=}&
     \sum_{\alpha^{A_t}\in \mathcal{A}_{t}} {P}(\alpha^{Q_{t+1}}|\underline{\alpha}^{Q_{t+1}},\alpha^{A_t}){P}(\alpha^{A_t}|{A}_{t},{U}_{t}),  \ t=0,\dots,T
\end{eqnarray*}
where
\begin{equation*}
{P}(\alpha^{Q_{t+1}}|\underline{\alpha}^{Q_{t+1}},\alpha^{A_t})\propto \prod_{m_{t}=1}^{M_t} {P}(\alpha_{tm_{t}}^{Q_{t+1}}|\alpha^{Q_{t+1}}_{t,-m_t},\underline{\alpha}^{Q_{t+1}},\alpha_{tm_{t}}^{A_t}),
\end{equation*}
\begin{equation*}
 {P}(\alpha_{tm_{t}}^{Q_{t+1}}=1|\alpha^{Q_{t+1}}_{t,-m_t},\underline{\alpha}^{Q_{t+1}},\alpha_{tm_{t}}^{A_t}=0)={P}(\alpha_{tm_{t}}^{Q_{t+1}}=0|\alpha^{Q_{t+1}}_{t,-m_t},\underline{\alpha}^{Q_{t+1}},\alpha_{tm_{t}}^{A_t}=0)=\frac{1}{2},
\end{equation*}
\begin{equation*}
  {P}(\alpha_{tm_{t}}^{Q_{t+1}}=1|\alpha^{Q_{t+1}}_{t,-m_t},\underline{\alpha}^{Q_{t+1}},\alpha_{tm_{t}}^{A_t}=1)=\frac{\omega_{tm_{t}}^{\alpha^{Q_{t+1}}}}{\omega_{tm_{t}}^{\alpha^{Q_{t+1}}}+1},
\end{equation*}
\begin{equation*}
{P}(\alpha_{tm_{t}}^{Q_{t+1}}=0|\alpha^{Q_{t+1}}_{t,-m_t},\underline{\alpha}^{Q_{t+1}}, \alpha_{tm_{t}}^{A_t}=1)=\frac{1}{\omega_{tm_{t}}^{\alpha^{Q_{t+1}}}+1}.
\end{equation*}
The rest of the presentation is exactly as in Section 4.2 of the main text.

\clearpage

\section{Additional results for the real-data illustration}

\begin{table}[h]
 \centering
 \caption{Baseline characteristics of the study population}  
 \label{tab:baseline_characteristics}
 \scalebox{0.8}{
   \begin{tabular}{lr}  
   \toprule
   \textbf{Characteristic} & \textbf{N = 14 020} \\  
   \midrule
   Age (mean (SD)) & 71.59 (5.37) \\  
   Prevalent diabetes, (yes, n($\%$)) & 2 625 (18.7) \\  
   Hypertension, (yes, n($\%$) & 3 515 (25.1) \\  
   Number of filled prescriptions in the year of statin initiation, (mean (SD)) & 4.71 (3.38) \\  
   Number of days of hospitalization in the year of statin initiation, (mean (SD)) & 0.24 (0.59) \\  
   Use of aspirin and other antiplatelet agents (yes, n($\%$)) & 615 (4.4) \\  
   Chronic kidney disease, (yes, n($\%$)) & 105 (0.7) \\  
   Blood pressure treatment, (yes, n($\%$)) & 8 490 (60.6) \\  
   CVD (yes, n(\%)) & 1 745 (12.4) \\  
   \bottomrule
   \end{tabular}}
   $\left. \right.$ \\
   \textit{Note:} CVD = Cardiovascular disease; SD = Standard deviation.   
\end{table} 

\begin{figure}[ht!]
	\centerline{\includegraphics[width=12in,width=\textwidth]{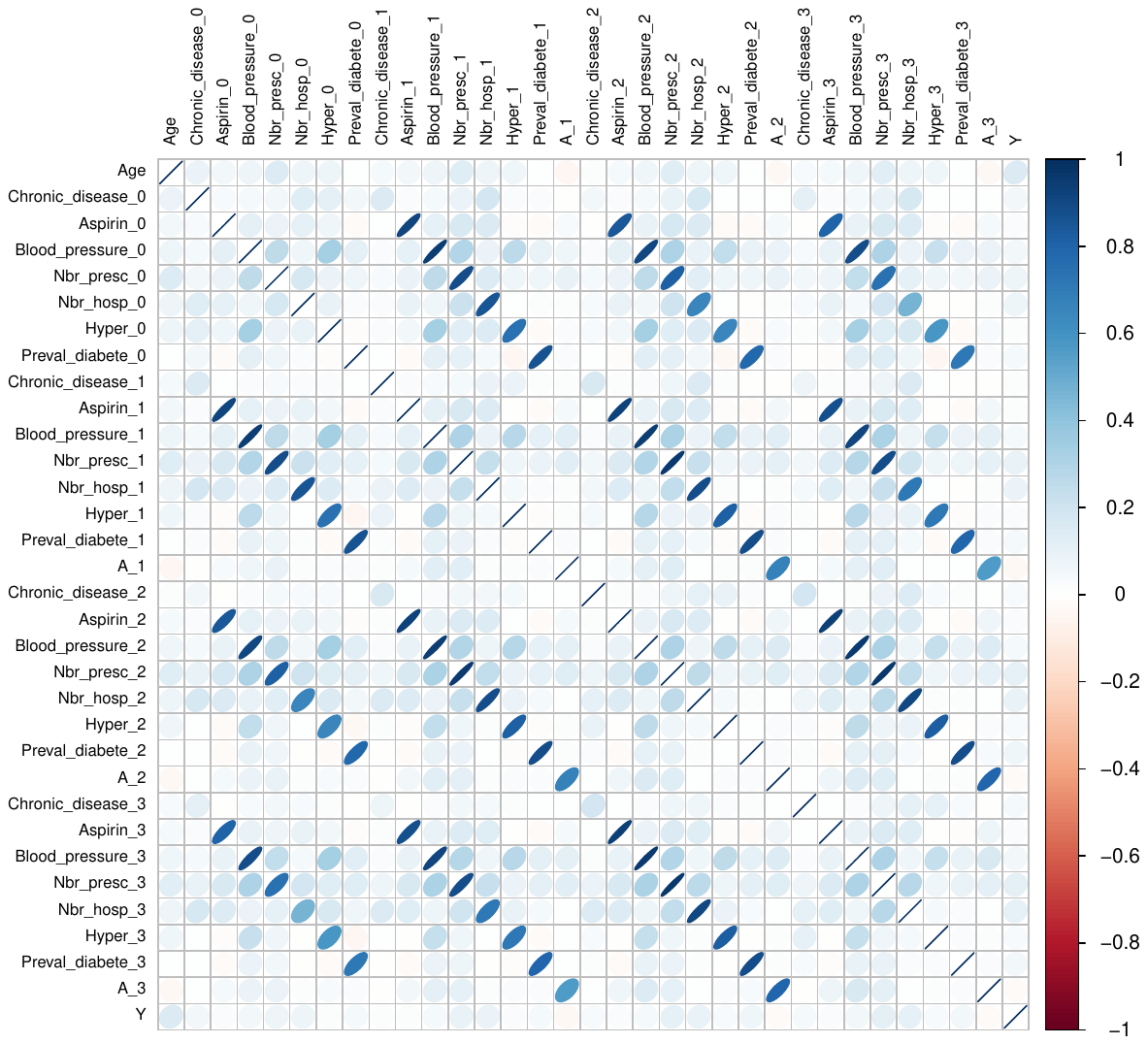}}
	\caption{Correlation between covariates at each time-point. Variables $A\_1$, $A\_2$ and $A\_3$ represent statin use at time $t = 1$,  $t = 2$ and $t = 3$ respectively.\label{figA17}}
\end{figure}

\end{document}